\documentclass[usenatbib]{mn2e}
\usepackage{aasmacros, amsmath, amssymb, color, graphicx}
\newcommand{\msol}{\ensuremath{~\mbox{M}_\odot}}

\newcommand{\mpc}{\ensuremath{~\mbox{Mpc}}}
\newcommand{\hmpc}{\ensuremath{\,h^{-1}\,\mbox{Mpc}}}

\newcommand{\hkpc}{\ensuremath{\,h^{-1}\,\mbox{kpc}}}
\newcommand{\gyr}{\ensuremath{~\mbox{Gyr}}}
\newcommand{\myr}{\ensuremath{~\mbox{Myr}}}

\newcommand{\yrinv}{\ensuremath{~\mbox{yr}^{-1}}}

\newcommand{\mstel}{\ensuremath{M_{\rm star}}}

\newcommand{\mhalo}{\ensuremath{M_{\rm halo}}}

\newcommand{\msubhalo}{\ensuremath{M_{\rm (sub)halo}}}
\newcommand{\mvir}{\ensuremath{M_{\rm vir}}}
\newcommand{\rvir}{\ensuremath{R_{\rm vir}}}

\newcommand{\mthm}{\ensuremath{M_{200{\rm m}}}}
\newcommand{\rthm}{\ensuremath{R_{200{\rm m}}}}

\newcommand{\fcenq}{\ensuremath{f^{\rm cen}_Q}}
\newcommand{\fcena}{\ensuremath{f^{\rm cen}_A}}
\newcommand{\fsatq}{\ensuremath{f^{\rm sat}_Q}}
\newcommand{\fsata}{\ensuremath{f^{\rm sat}_A}}

\newcommand{\fsatqexcess}{\ensuremath{f^{\rm sat}_{Q\,\rm excess}}}

\newcommand{\sfr}{\ensuremath{\mbox{SFR}}}

\newcommand{\ssfr}{\ensuremath{\mbox{SSFR}}}
\newcommand{\dnfk}{\ensuremath{\mbox{D}_n4000}}

\newcommand{\halpha}{\ensuremath{\mbox{H}\alpha}}

\newcommand{\zinf}{\ensuremath{z_{\rm inf}}}

\defcitealias{TinWetCon11}{Paper~I}
\defcitealias{WetTinCon11}{Paper~II}
\defcitealias{WetTinCon12a}{Paper~III}
\defcitealias{WetTinCon12b}{Paper~IV}
\defcitealias{Tin12}{Tinker et al., in prep.}

\begin{document}
\title[Galaxy evolution in groups and clusters]{Galaxy evolution in groups and clusters: star formation rates, red sequence fractions, and the persistent bimodality}
\author[Wetzel, Tinker \& Conroy]{Andrew R. Wetzel${}^1$, Jeremy L. Tinker${}^2$ and Charlie Conroy${}^3$\\
$^{1}$Department of Astronomy, Yale University, New Haven, CT 06520, USA\\
$^{2}$Center for Cosmology and Particle Physics, Department of Physics, New York University, New York, NY 10013, USA\\
$^{3}$Harvard-Smithsonian Center for Astrophysics, Cambridge, MA 02138, USA
}
\date{July 2011}
\pagerange{\pageref{firstpage}--\pageref{lastpage}} \pubyear{2011}
\maketitle
\label{firstpage}

%% ABSTRACT %%%%%%%%%%%%%%%%%%%%%%%%%%%%%%%%%%%%%%%%%%%%%%%%%%%%%%%%%%%%%%%%%%%%%%%%%%%%%%
\begin{abstract}
Using galaxy group/cluster catalogs created from the Sloan Digital Sky Survey Data Release 7, we examine in detail the specific star formation rate (SSFR) distribution of satellite galaxies and its dependence on stellar mass, host halo mass, and halo-centric radius.
All galaxies, regardless of central-satellite designation, exhibit a similar bimodal SSFR distribution, with a strong break at $\ssfr \approx 10^{-11}\yrinv$ and the same high SSFR peak; in no regime is there ever an excess of galaxies in the `green valley'.
Satellite galaxies are simply more likely to lie on the quenched (`red sequence') side of the SSFR distribution.
Furthermore, the satellite quenched fraction excess above the field galaxy value is nearly independent of galaxy stellar mass.
An enhanced quenched fraction for satellites persists in groups with halo masses down to $3 \times 10^{11}\msol$ and increases strongly with halo mass and toward halo center.
We find no detectable quenching enhancement for galaxies beyond $\sim2\,\rvir$ around massive clusters once these galaxies have been decomposed into centrals and satellites.
These trends imply that (1) galaxies experience no significant environmental effects until they cross within $\sim\rvir$ of a more massive host halo, (2) after this, star formation in active satellites continues to evolve in the same manner as active central galaxies for several Gyrs, and (3) once begun, satellite star formation quenching occurs rapidly.
These results place strong constraints on satellite-specific quenching mechanisms, as we will discuss further in companion papers.
\end{abstract}


%% INTRODUCTION %%%%%%%%%%%%%%%%%%%%%%%%%%%%%%%%%%%%%%%%%%%%%%%%%%%%%%%%%%%%%%%%%%%%%%%%%%
\section{Introduction}

Galaxy properties depend on their local environment.
Galaxies in denser regions, such as groups and clusters, exhibit attenuated star formation rates (SFR), and therefore significantly higher red sequence fractions, as well as more elliptical morphologies as compared with galaxies in less dense environments \citep{Oem74, DavGel76, Dre80, PosGel84}.
With the advent of voluminous galaxy surveys such as the Sloan Digital Sky Survey \citep [SDSS;][]{SDSS}, numerous works have quantified the correlations between these galaxies properties and their environment in considerable detail \citep[e.g.,][]{HogBlaBri04, BalBalNic04, KauWhiHec04, BlaEisHog05}, and more recent works have shown that similar environmental dependence persists at least out to $z \approx 1$ \citep{CucIvoMar06, CooNewCoi07, PenLilKov10}.

There are two particularly interesting features of the dependence of these properties on environment.
First, environmental impact on SFR (or color) is stronger than on morphology \citep{KauWhiHec04, BlaEisHog05, BalLovBru08, BamNicBal09}: at fixed morphology, SFR still exhibits strong changes with environment, but at fixed SFR, there is almost no dependence of morphology on environment.
Second, SFR/color and morphology primarily depend on small-scale ($\lesssim 1\mpc$) environment, with little-to-no additional dependence on environment measured on larger scales \citep{HogBlaBri04, KauWhiHec04, BlaEisHog05}.
Several works have extended this result through the use of galaxy group catalogs, showing that galaxy color and star formation history most directly depend on the properties of their host dark matter halo \citep{BlaBer07, WilZibBud10, TinWetCon11}.
Thus, understanding galaxy star formation requires identifying and understanding a galaxy's host halo properties.

The idea that environmental dependence primarily manifests itself via host halo properties is motivated by the fact that the region within a halo is physically distinct from the field environment.
The halo virial radius corresponds to a physical transition between infall and virialized motions, capable of supporting strong shock fronts, at least in halos more massive than $\sim10^{12}\msol$ \citep[e.g.,][]{DekBir06}.
Within the virial radius, high densities lead to strong tidal forces, and accreted gas is thermalized to high temperature and pressure.
`Satellite' galaxies that fall into more massive halos thus experience tidal and ram-pressure stripping, heating, and are unable to accrete gas as efficiently as in the field.
Thus, within groups and clusters, it is important to distinguish satellite galaxies from the galaxy at the minimum of the host halo potential well (the `central' galaxy), which occupies a special dynamical location and is typically the oldest and most massive in the halo.

In this paper, we focus on the SFRs of satellites galaxies, how they depend on their host halo properties, and how they compare with central galaxies, which have not (yet) fallen into a more massive halo.
Understanding satellite star formation is important not only for elucidating the physical processes that occur within groups and clusters, but also for a comprehensive understanding of galaxy evolution.
For example, satellites are responsible for more than a 1/3 of the build-up of the red sequence population \citep{vdBAquYan08, TinWet10}.
In addition, many methods of optically identifying galaxy groups/clusters rely on selecting red sequence galaxies \citep[e.g.,][]{KoeMcKAnn07a} because they are more prominent within groups/clusters, making it important to characterize in detail how the satellite red sequence fraction depends on halo mass and radius.

Even though it has long been known that galaxies in groups/clusters are significantly more likely to have attenuated SFRs \citep{DreGun83, BalMorYee97, PogSmaDre99}, the mechanism(s) that dominates this transformation remains unclear, though several have been proposed.
As a satellite galaxy orbits through its host halo's hot, virialized gas, ram-pressure can strip cold gas directly from the disc \citep{GunGot72, AbaMooBow99}, causing a rapid ($\lesssim500\myr$) quenching of star formation.
There are striking, direct examples of ram-pressure stripping occurring for satellites in nearby clusters \citep[e.g.,][]{ChuvGoKen09}, some of which are even near the virial radius.
While this suggests that ram-pressure stripping might be responsible for satellite quenching in massive clusters, even at large radii, it remains unclear whether ram-pressure can act efficiently in lower mass groups, particularly if virial shock fronts are not easily supported.

Other mechanisms affect gas accretion, both into the subhalo that surrounds a satellite and cooling onto the disc.
Referred to as `strangulation' or `starvation', this more gradual process involves the lack of accretion of new hot gas, as well as loss of existing extended hot gas via heating/evaporation, tidal stripping, or ram-pressure stripping \citep{LarTinCal80, BalNavMor00, KawMul07, McCFreFon08}, causing SFR in the disc to decline on a cold gas consumption timescale.
X-ray observations show direct evidence for the truncation of hot gas halos around satellite galaxies in groups and clusters \citep{SunJonFor07, JelBinMul08}.

Finally, galaxies in groups and clusters can interact with each other gravitationally.
`Harassment' involves high-speed, nearby encounters that tidally heat satellites \citep{FarSha81, MooLakKat98}.
More violently, satellites can merge with one another \citep{MakHut97}, a process that may not be uncommon given that satellite orbits are strongly correlated \citep{AngLacBau09, WetCohWhi09a, WetCohWhi09b, WhiCohSmi10, Coh12}.
These dynamical processes potentially could induce rapid gas consumption and thus quenching, in addition to driving morphological transformation.

To understand which of the above physical process(es) dominates, it is crucial to have a detailed understanding of how satellite SFR depends on its host halo's mass and radius within the halo, including whether there is a minimum halo mass for affecting satellite SFR.
The above physical processes have different predictions for these dependences.
However, the extent to which a satellite's SFR depends on its host halo mass, and how this compares with intrinsic dependence on its stellar mass, remains in debate.
Several recent works using SDSS group/cluster catalogs found evidence that satellites are more likely to be quenched/red in more massive halos:
\citet{WeivdBYan06a} examining both SFR and color using the group catalog finder of \citet{YanMovdB05a}; \citet{BlaBer07} examining color using their own group catalog; \citet{KimSomYi09} examining UV-based SFR using the group catalog of \citet{YanMovdB07}; and \citet{HanSheWec09} examining color in the MaxBCG catalog as a function of cluster richness.
However, other works conclude that the satellite SFR/color dependence on halo mass is non-existent or at least much smaller than the intrinsic dependence on stellar mass:
\citet{FinBalZar08} examining SFR in the C4 cluster catalog as a function of cluster velocity dispersion; and \citet{vdBPasYan08}b and \citet{PasvdBMo09} examining color and SFR, respectively, using the group catalog of \citet{YanMovdB07}.
Similarly, many recent works found that satellites are more likely to be red \citep{DePColPea04, BlaBer07, HanSheWec09} and have lower SFRs \citep{BalNavMor00, EllLinYee01, WeivdBYan06a, vdLWilKau10, PreBalJam11} toward group/cluster center, including weak trends for groups at $z \approx 1$ \citep{GerNewFab07}.
Though again, there is debate about the steepness of radial gradients, the extent to which they depend on halo mass, and whether they are a byproduct of satellite mass segregation \citep{vdBPasYan08, PasvdBMo09}.

There are several possible reasons for these disagreements.
First, one must compare galaxies at fixed stellar mass because SFR and color depend strongly on mass \citep{KauHecWhi03}, and more massive galaxies are more common in denser environments \citep{HogBlaEis03}.
The environmental trends noted in many previous works were largely a reflection of these intrinsic stellar mass dependencies.
Second, some SFR metrics are more susceptible to systematic biases than others, and in particular, simple color cuts can overestimate the quenched fraction because of dust reddening \citep[e.g.,][]{MalBerBla09}.
Third, in order to elucidate these trends, it is important to study satellite properties over a wide range of both satellite and host halo masses.

Finally, to fully understand the physical mechanisms operating on satellites, it is crucial to go beyond simple SFR/color cuts and understand the nature of the full SFR distribution.
\citet{BalBalNic04} examined the $u-r$ color distribution of galaxies in SDSS as a function of projected galaxy environment (defined by fifth nearest neighbor) and found that the color bimodality extended to all environments, and that the colors of blue galaxies do not depend significantly on environment.
Using several different SDSS group catalogs, \citet{Ski09} showed that the $g-r$ color distribution of satellites persists across group varying richness.
More recently, \citet{McGBalWil11} found similar trends examining the SFR distributions of all galaxies in groups and in the field both in SDSS and at $z \approx 0.4$.

In this work, we explore the satellite galaxy SFR distribution in detail, including its dependence on host halo mass and halo-centric radius, to robustly characterize the properties of satellite star formation evolution and quenching in groups and clusters.
To observationally determine the host halo masses of SDSS galaxies, we construct galaxy group catalogs using a variant of the \cite{YanMovdB05a} group-finding algorithm, which also allows us to examine how the SFRs of satellites differ from central galaxies of similar mass.
Our galaxy sample and group finder allows us to explore a considerable range of halo and satellite masses to determine the regimes in which halo-specific processes are important.
We examine galaxies selected in narrow bins of stellar mass to isolate host halo dependence from intrinsic stellar mass dependence, and we use spectroscopically-derived SFR measurements which are free from dust reddening effects.

This paper represents the second in a series of four.
In \citet{TinWetCon11}, hereafter referred to as \citetalias{TinWetCon11}, we presented our method for identifying galaxy groups in SDSS.
Using the $4000$\AA\ break, $\dnfk$, as a measure of recent star formation history, we showed that central and satellite galaxy star formation is nearly independent of large-scale environmental density beyond its dark matter host halo.
In this work, we explore the full SFR distribution of satellite galaxies\footnote{
We have examined all trends in this paper selecting galaxies on both $M_r$ and $\mstel$ as well as examining both $\dnfk$ and SSFR, and while these lead to slight quantitative differences, they do not change any of our results qualitatively.}
and its dependence on stellar mass, halo mass, and halo-centric radius to fully characterize the local density dependence of satellite star formation.
In \citet{WetTinCon12a}, hereafter referred to as \citetalias{WetTinCon12a}, we will combine the results of this paper with a high-resolution cosmological simulation that tracks satellite infall times in order to examine satellite star formation histories and quenching timescales in detail.
Finally, in \citet{WetTinCon12b}, hereafter referred to as \citetalias{WetTinCon12b}, with the results of this paper we will use satellite orbital histories from simulation to test and constrain the physical mechanisms of satellite-specific quenching.

For all calculations we use a flat, $\Lambda$CDM cosmology of $\Omega_{\rm m}=0.27$, $\Omega_{\rm b}=0.045$, $h=0.7$, $n_{\rm s}=0.95$ and $\sigma_8=0.82$, consistent with a wide array of observations \citep[see, e.g.,][and references therein]{KomSmiDun11}.

%% METHODS %%%%%%%%%%%%%%%%%%%%%%%%%%%%%%%%%%%%%%%%%%%%%%%%%%%%%%%%%%%%%%%%%%%%%%%%%%%%%%%
\section{Methods} \label{sec:method}

The details of our galaxy sample and group finding algorithm are discussed extensively in \citetalias{TinWetCon11}.
Here, we highlight the aspects relevant to this work.

\subsection{Galaxy catalog} \label{sec:galaxy_catalog}

To construct our galaxy sample, we use the New York University Value-Added Galaxy Catalog \citep{BlaSchStr05} based on SDSS Data Release 7 \citep{AbaAdeAgu09}.
This spectroscopic catalog ensures that accurate galaxy redshifts, stellar masses, and SFRs can be obtained.
Galaxy stellar masses are based on the {\tt kcorrect} code of \citet{BlaRow07}, assuming a \citet{Cha03} initial mass function.

To construct stellar mass-limited samples, we first construct two volume-limited samples of all galaxies with $M_r < -18$ and $M_r < -19$, which contain galaxies out to $z = 0.04$ and $z = 0.06$.
Within these magnitude limits, we identify minimum complete stellar mass limits of $5 \times 10^{9}\msol$ and $1.3 \times 10^{10}\msol$, respectively.
Combining these samples leads to an overall median redshift of $z = 0.045$.

For a galaxy star formation metric, we use specific star formation rate (SSFR), where $\ssfr = \sfr/\mstel$.
These values are based on the current release\footnote{
\tt http://www.mpa-garching.mpg.de/SDSS/DR7/} 
of the spectral reductions of \citet{BriChaWhi04}, with updated prescriptions for active galactic nuclei (AGN) contamination and fiber aperture corrections following \citet{SalRicCha07}.
These SSFRs are primarily derived from emission lines (mostly $\halpha$), but in the cases of strong AGN contamination or no measurable emission lines, the SSFRs are inferred from $\dnfk$.
Roughly, $\ssfr \gtrsim 10^{-11}\yrinv$ are based almost entirely on $\halpha$, $10^{-12} \lesssim \ssfr \lesssim 10^{-11}\yrinv$ are based on a combination of emission lines, and $\ssfr \lesssim 10^{-12}\yrinv$ are based almost entirely on $\dnfk$ and should be considered upper limits to the true value.

These SSFRs have also been corrected for aperture bias because the SDSS spectral fiber size of 3 arcsec ($1.5\hkpc$ at our median redshift) can be smaller than a galaxy's effective radius.
Specifically, \citet{BriChaWhi04} measured the distribution of SSFRs at a given $g-r$ and $r-i$ color within the fiber for all galaxies and applied this empirical calibration to the photometry outside the fiber.
Their updated, aperture-corrected SSFRs yield good agreement with full-galaxy aperture, $UV$-based SSFR measurements of \citet{SalRicCha07}.

Galaxy pairs closer than 55 arcsec ($30\hkpc$ at our median redshift) are too close to both receive spectral fibers (`fiber collisions'), so 5\% of the galaxies in our sample do not have spectra.
These galaxies are assigned the same redshift as their nearest non-collided galaxy, and we assign SSFR by sampling randomly from the distributions of non-collided galaxies with the same $r$-band magnitude and $g-r$ color within the volume-limited sample.
We have analyzed all of our results both using and excluding fiber-collided galaxies, and both produce consistent results, though to be safe in examining trends with radius we only examine radial bins where the fiber-collided fraction is less than 10\%.

We emphasize that the use of spectroscopically-derived SSFRs is critical for our analysis because dust reddening can severely overestimate the quenched population when using red/blue color cuts, particularly at lower mass where gas and dust fractions are higher.
For our lowest mass galaxies ($\mstel < 10^{10}\msol$) 37\% of red ($g-r > 0.76$) galaxies are in fact active ($\ssfr > 10^{-11}\yrinv$), and 21\% of active galaxies appear red (see Fig.~2 in \citetalias{TinWetCon11}).
We have confirmed that this population of active but red galaxies is caused primarily by dusty spirals by applying Galaxy Zoo morphological classifications \citep{LinSchBam11} to this population: $\sim70\%$ are identified as spirals, and $\sim50\%$ are edge-on spirals \citep[see also][]{MasNicBam10}.

\subsection{Galaxy group catalog} \label{sec:group_catalog}

Motivated by the paradigm that all galaxies reside in host dark matter halos, we identify the halo properties of galaxies by using a modified implementation of the group-finding algorithm detailed in \citet{YanMovdB05a,YanMovdB07}.
We define a galaxy group as a set of galaxies that occupy the same host dark matter halo, and we will use `group' and `halo' interchangeably henceforth.

For our group catalog, we define dark matter halos such that the mean matter density interior to the virial radius is 200 times the mean background matter density: $\mthm = 200 \bar{\rho}_{\rm m} \frac{4}{3} \pi \rthm^3$.
Using the host halo mass function from \citet{TinKraKly08} combined with the satellite subhalo mass function from \citet{TinWet10}, we first assign a tentative (sub)halo mass to each galaxy by matching the abundance of objects above a given mass: $n(>\msubhalo) = n(>\mstel)$.
Each galaxy then has an associated (sub)halo virial radius and velocity dispersion.
Starting with the most massive galaxy in the sample, we assign to each nearby, lower mass galaxy a probability of being a satellite member of its group using a matched filter in both radius and line-of-sight velocity difference.
We repeat this process for less massive galaxies on down the mass function, skipping those that have already been assigned as satellites in a group, until each galaxy is assigned to a (tentative) group.
Using now the total stellar mass of each group, we update the halo mass of each group based on abundance matching, but now using only the host halo mass function (ignoring subhalos): $n(>M_{\rm halo}) = n(>M_{\rm star,group})$.
With these updated halo masses, we re-assign satellite membership probabilities, we update the halo masses again, and we iterate until convergence.
Every group thus contains one `central' galaxy, which by definition is the most massive, and a group can contain zero, one, or many satellites.

We construct groups at $z < 0.04$ using the $\mstel > 10^{9.7}\msol$ sample and at $z = 0.04 - 0.06$ using the $\mstel > 10^{10.1}\msol$ sample.
This leads to $\sim19000$, 2200, 160 groups with $\mthm > 10^{12}$, $10^{13}$, $10^{14}\msol$, respectively, up to the most massive group at 
$10^{15}\msol$ hosting 269 satellites.
Our catalog allows us to examine satellites with good statistics in the mass range $\mstel = 5 \times 10^{9} - 2 \times 10^{11}\msol$.

We define a group's center by the location of its central (most massive) galaxy.
However, in reality the most massive galaxy is not always at the minimum of a host halo's potential well, for two reasons.
First, the galaxy at the minimum of the potential might not be the most massive in the halo.
Using the \citet{YanMovdB07} group catalog, \citet{SkivdBYan11} found evidence that up to 40\% of massive ($>10^{14}\msol$) halos have galaxies closest to halo center that are not the brightest, with a decreasing fraction at lower halo mass.
However, these are typically cases where two galaxies in a halo have similarly high luminosities, and in this regime galaxies exhibit highly quenched SSFRs almost regardless of central vs. satellite demarcation, so we do not expect that this effect will bias our satellite SSFR results significantly.
Second, the (true) central galaxy can be offset from the minimum of the halo potential as a result of dynamics processes such as merging.
For example, for rich clusters in the SDSS MaxBCG catalog with X-ray detection, the median offset between the brightest cluster galaxy and X-ray center is $58\hkpc$ \citep{KoeMcKAnn07b}.
To test the impact of both of these possible biases, we have examined our radius trends centering on both the most massive galaxy and the luminosity-weighted center of the group, and we find no significant difference in our results.

Because of projection effects and redshift-space distortions, some central galaxies are inevitably misassigned as `satellites' of higher mass halos (reducing group purity), and conversely some satellites are misassigned as `central' galaxies of lower mass halos (reducing group completeness).
As detailed in \citetalias{TinWetCon11}, we find that an average of $\sim10\%$ of galaxies have been misassigned, with a stronger effect for more massive satellites in lower mass halos \citep[see also][]{YanMovdB07}.
Because central galaxies exhibit a lower quenched fraction than satellites, and because they outnumber satellites, the primary resultant biasing is an underestimation of the quenched fraction for satellites.
This effect also causes an overestimation of the quenched fraction for central galaxies, but because central galaxies outnumber satellites this effect is relatively small.
In this work, we do not attempt to correct our results for these biases given their non-trivial dependence on halo mass and radius.

Finally, we have checked our results against a SDSS DR7 version of the \citet{YanMovdB07} group catalog (kindly provided by Frank van den Bosch), incorporating the same {\tt kcorrect} stellar masses and SSFR measurements we use here but retaining differences in assumed cosmology, halo virial definitions, and group-making methodology.
All of the results are consistent within errors.

%% MASS %%%%%%%%%%%%%%%%%%%%%%%%%%%%%%%%%%%%%%%%%%%%%%%%%%%%%%%%%%%%%%%%%%%%%%%%%%%%%%%%%%
\section{Mass dependence} \label{sec:mass}

\subsection{Star formation rate distribution}

\begin{figure}
\centering
\includegraphics[width=0.99\columnwidth]{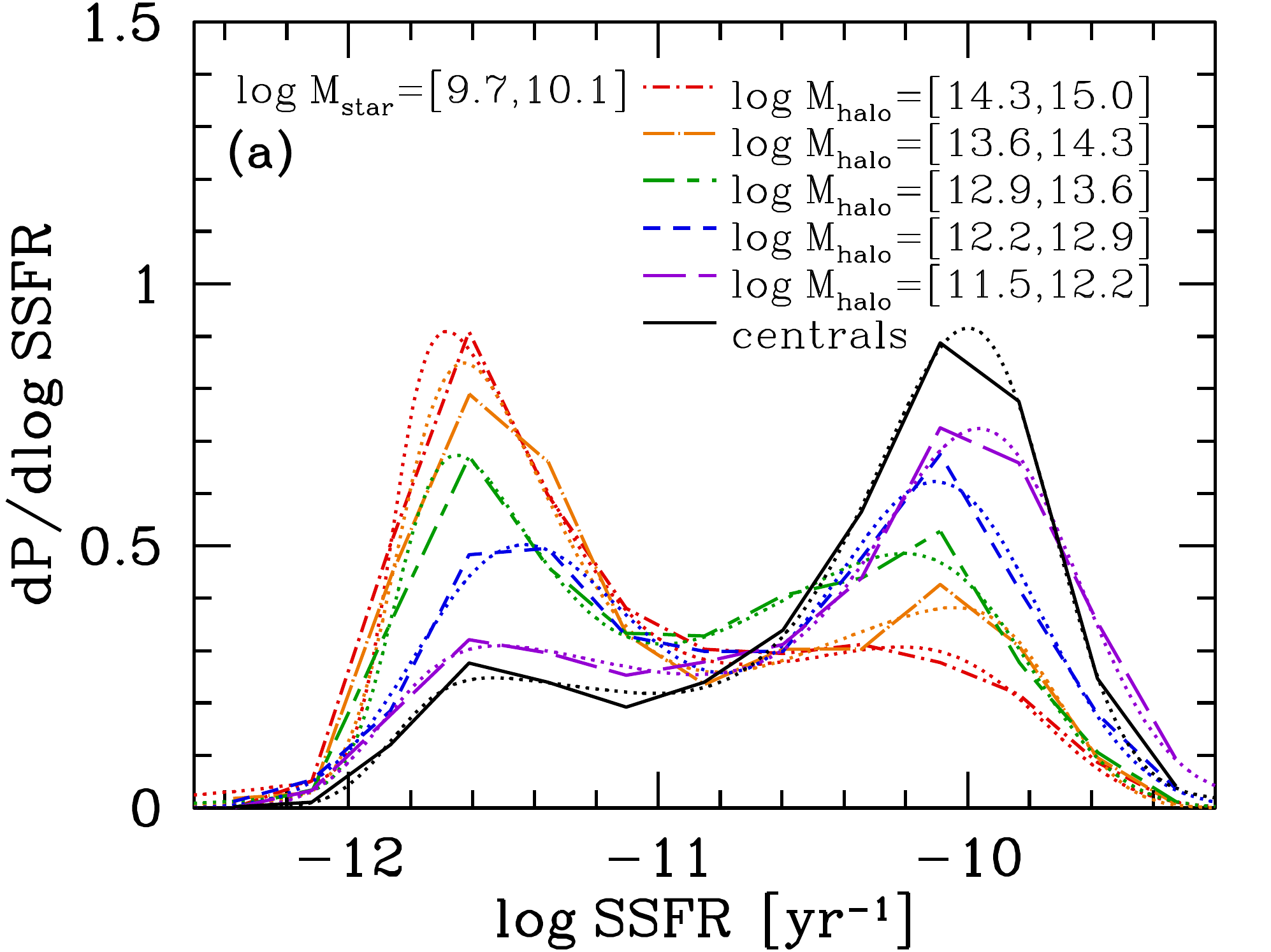}
\includegraphics[width=0.99\columnwidth]{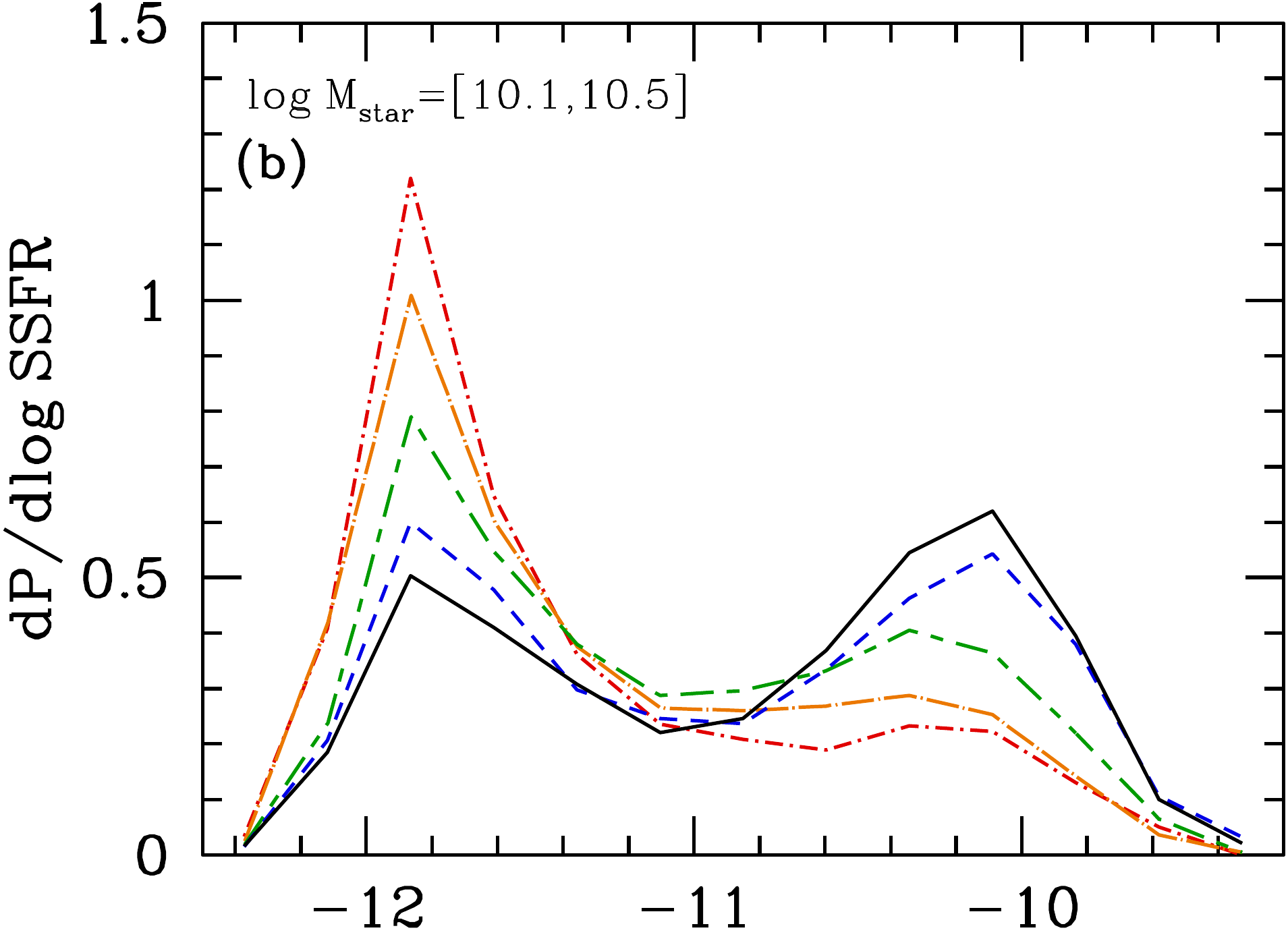}
\includegraphics[width=0.99\columnwidth]{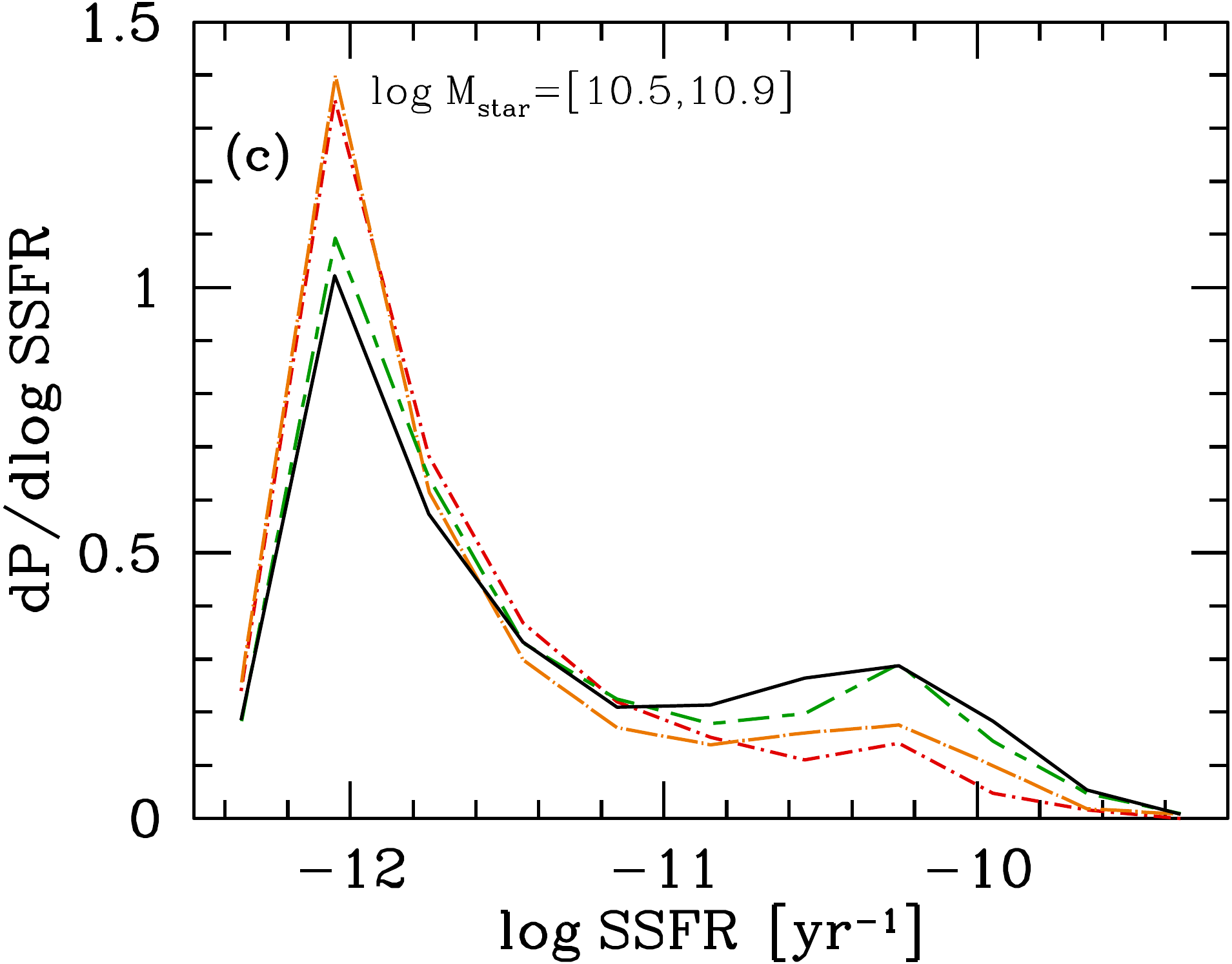}
\caption{
Specific star formation rate (SSFR) distribution for central galaxies (solid) and satellites in bins of host halo mass (dashed and dot-dashed).
All galaxies in all halo masses we probe exhibit similar bimodal SSFR distributions.
Satellites in more massive halos are more likely to be quenched, but the SSFR values of the high SSFR peak, the break at $\ssfr \approx 10^{-11}\yrinv$, and the fraction of galaxies at the break remain unchanged.
Dotted curves in (a) show example fits to a skewed double Gaussian.
} \label{fig:ssfr_distr-m}
\end{figure}

We begin by exploring the galaxy SSFR distribution and its dependence on galaxy and host halo mass.
Fig.~\ref{fig:ssfr_distr-m} shows the distribution of SSFRs for galaxies decomposed into centrals and satellites.
Each panel shows a different stellar mass bin, and satellites are divided into halo mass bins.
Examining first the central galaxy distribution, at all stellar masses it exhibits a clear bimodality with a break at $10^{-11}\yrinv$.
More massive central galaxies have a decreased likelihood of high SSFRs and an enhanced likelihood of low SSFRs, but the break location remains fixed.
This SSFR bimodality and fixed break value have been noted by several authors \citep[e.g.,][]{BriChaWhi04, KauWhiHec04}.

Satellites exhibit a clear deficit/excess at high/low SSFR as compared with central galaxies of the same stellar mass.
This difference is smaller in lower mass halos, but we emphasize that Fig.~\ref{fig:ssfr_distr-m}a shows that a clear difference persists even at $\mhalo \lesssim 10^{12}\msol$, and that this is a lower limit to the actual difference because we have not corrected for purity and completeness effects.
Thus, \textit{we find no evidence for a minimum halo mass for affecting satellite star formation.}
In more massive halos, the relative fraction of high SSFR galaxies decreases monotonically with halo mass across all halo masses we probe.
Remarkably, though, the SSFR values where the satellite bimodality break and active peak occur do not change, nor does the fraction of satellites lying at the $10^{-11}\yrinv$ bimodality break (the `green valley').

\begin{figure}
\centering
\includegraphics[width=0.99\columnwidth]{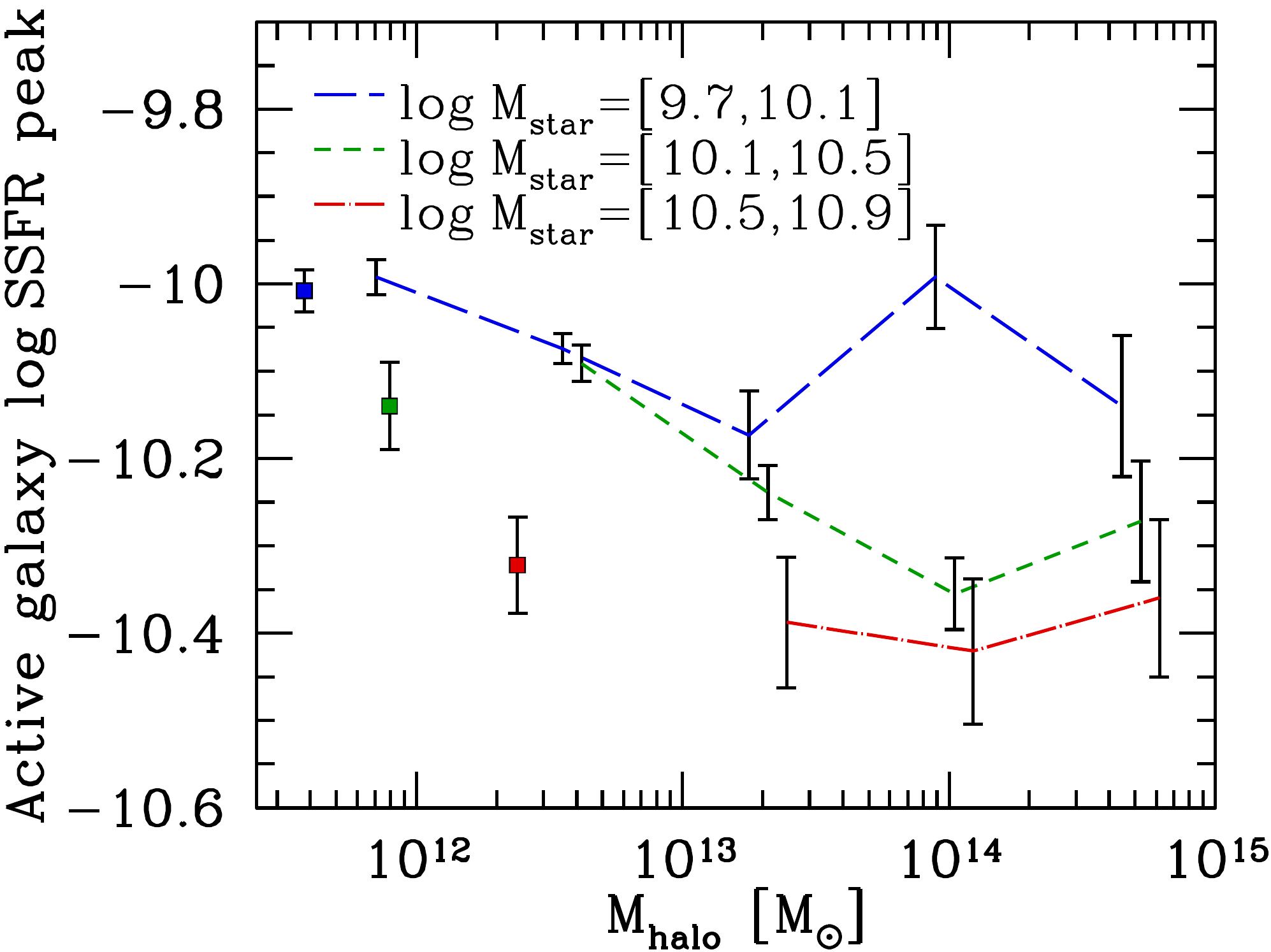}
\caption{
Active galaxy SSFR distribution peak vs. host halo mass, in bins of galaxy stellar mass, as determined by fitting a skewed, double Gaussian to each full SSFR distribution (see Fig.~\ref{fig:ssfr_distr-m}a for example).
Points show central galaxies, plotted at the median host halo mass of each sample, and curves show satellites.
While more massive galaxies show a trend toward lower SSFR peak values, there is no significant dependence on host halo mass.
} \label{fig:ssfr_peak-mhalo}
\end{figure}

We examine the invariance of the high SSFR peak more quantitatively by fitting a skewed, double Gaussian to the full SSFR distributions in Fig.~\ref{fig:ssfr_distr-m}, with example fits shown by dotted curves.
Fig.~\ref{fig:ssfr_peak-mhalo} shows the peak of the distribution for high SSFR (`active') galaxies as a function of host halo mass.
(Note that $\ssfr = 10^{-10.1}\yrinv$ corresponds to a rate that doubles mass in a Hubble time.)
Square points at left show central galaxies, plotted at the median halo mass corresponding to the stellar mass bin (recall that central galaxies have a tight relation between stellar and halo mass), while curves show satellites.
More massive galaxies exhibit systematically lower SSFR peak values, but within a given stellar mass bin there is no significant change with halo mass.
Thus, \textit{the SFRs of galaxies that remain active are not affected after falling into a host halo.}
We discuss the significance of these SSFR bimodality results further in \S\ref{sec:bimodality}.

We do not examine the peak at low SSFR, as it is partially an artifact of the spectral reductions of \citet{BriChaWhi04}, where low SSFR galaxies with no detectable emission lines are assigned SSFRs based on $\dnfk$, which effectively leads to an upper limit value of $\ssfr \approx 10^{-12}\yrinv$.
Thus, while the rise of the distribution below $10^{-11}\yrinv$ is real, the strong sharpness of the peak near $10^{-12}\yrinv$ is largely artificial, and in reality the distribution exhibits a tail to much lower SSFR \citep{SalRicCha07}.

Given the invariance of the value of the SSFR bimodality break across galaxy mass, host halo mass, and satellite-central demarcation, we divide galaxies into those with $\ssfr > 10^{-11}\yrinv$, referred to as `active', and those with $\ssfr < 10^{-11}\yrinv$, referred to as `quenched'.
With this active/quenched demarcation, we will examine the fraction of galaxies at a particular stellar mass that have quenched SSFRs, that is, the `quenched fraction'.
This quantity represents a more physically meaningful version of the `red sequence fraction', and the two are effectively synonymous in the limit of no dust reddening.

\subsection{Quenched fraction}

\begin{figure}
\centering
\includegraphics[width=0.99\columnwidth]{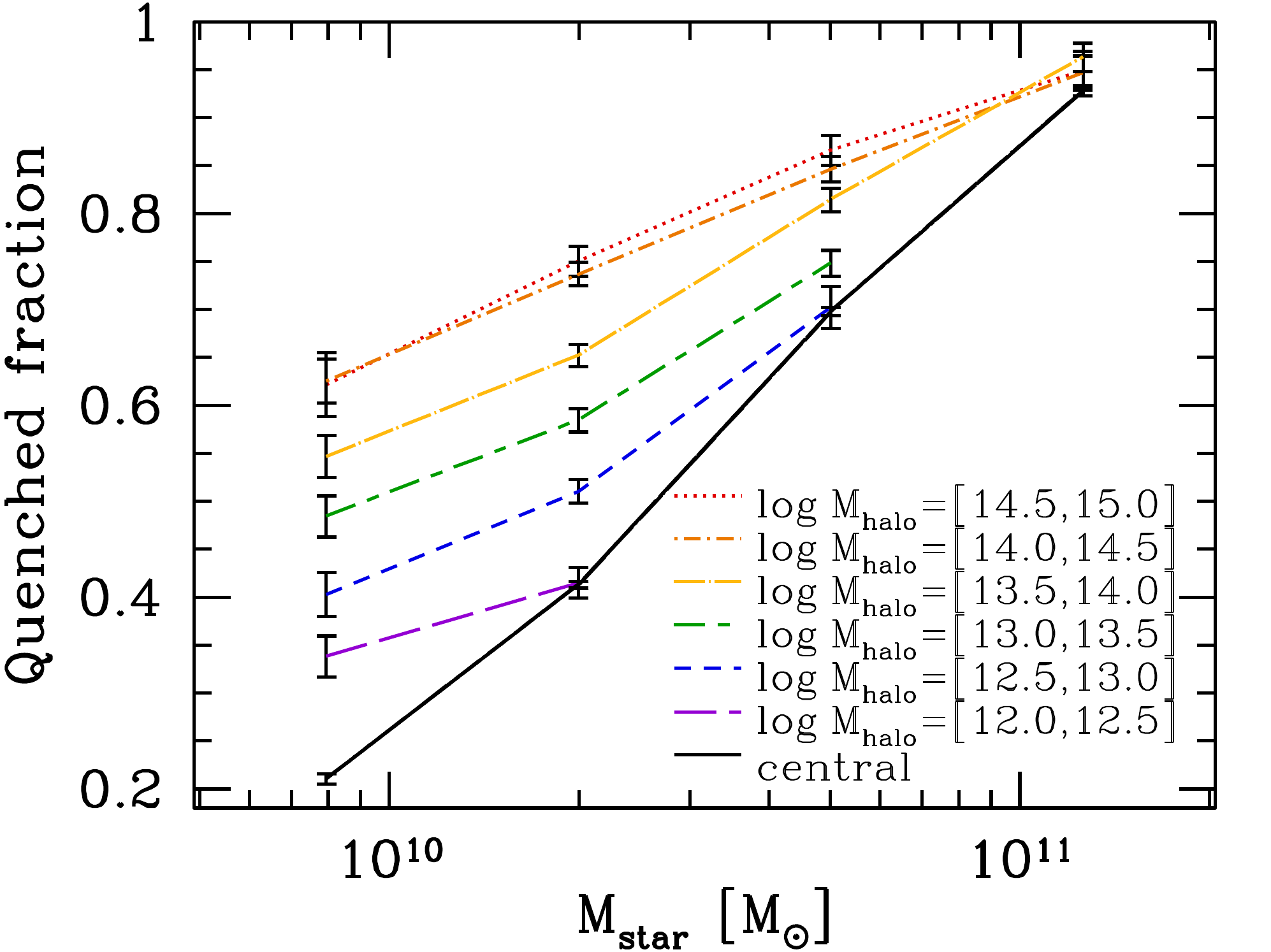}
\includegraphics[width=0.99\columnwidth]{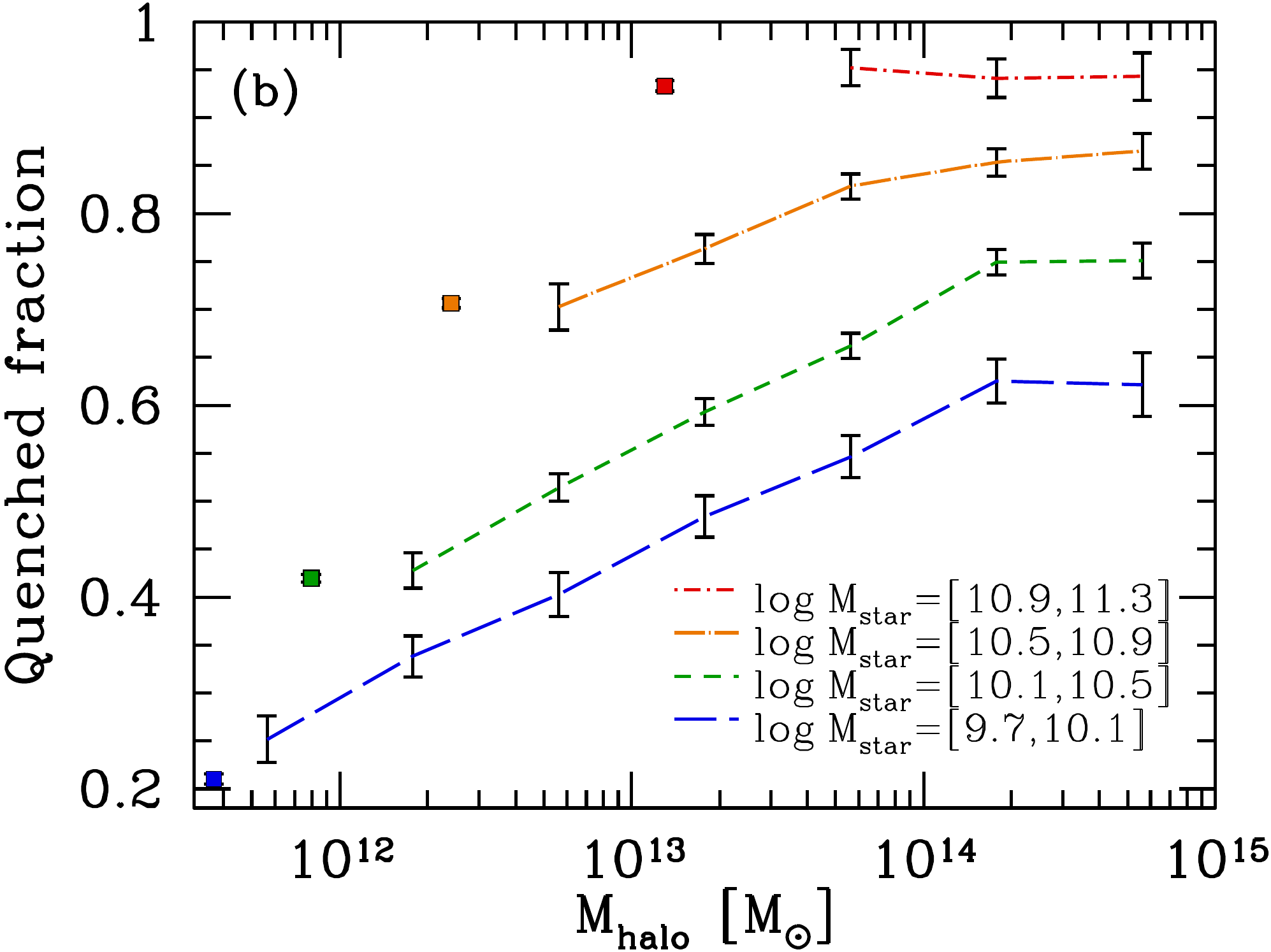}
\includegraphics[width=0.99\columnwidth]{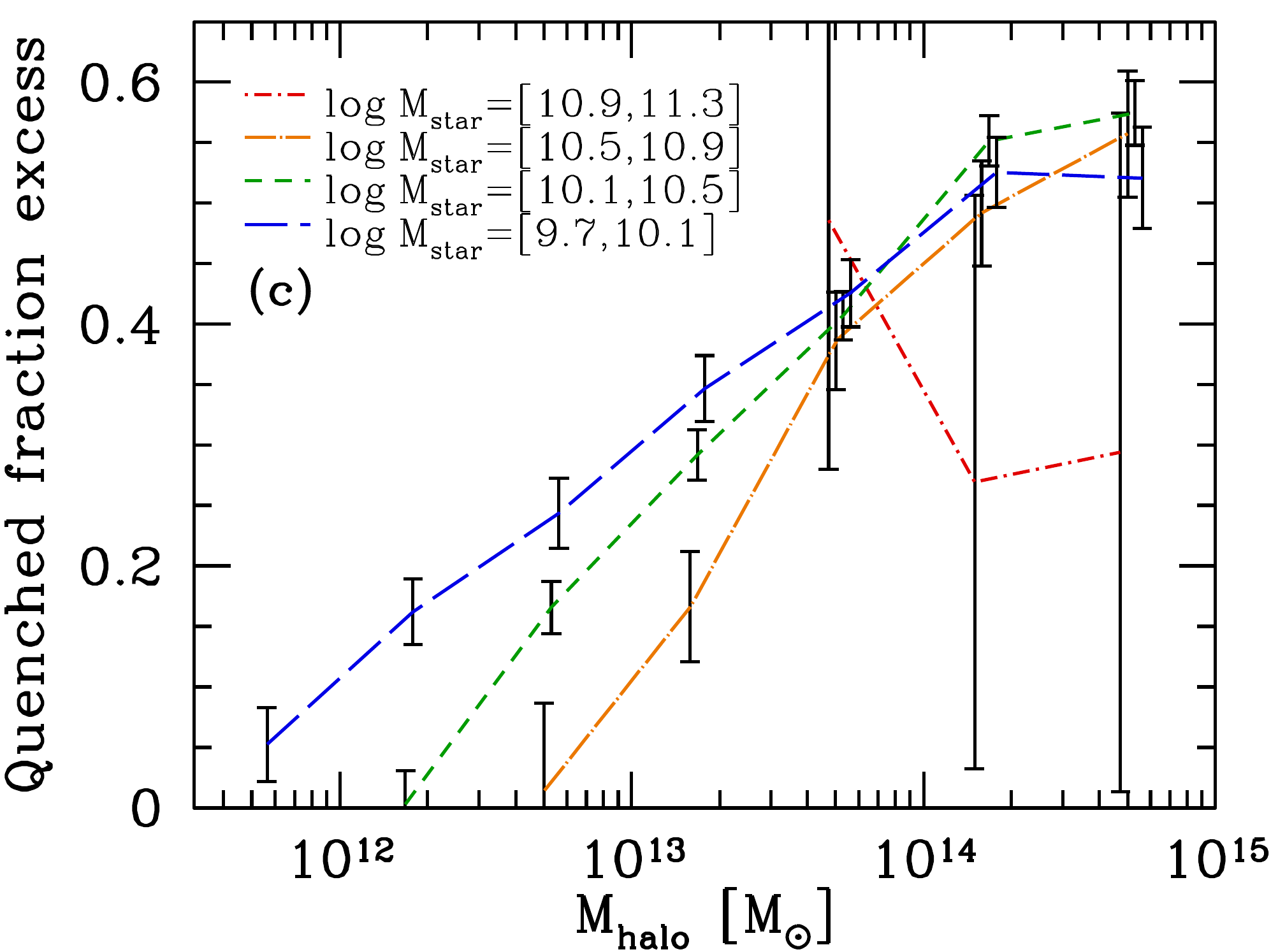}
\caption{
\textbf{(a)}: Galaxy quenched fraction ($\ssfr < 10^{-11}\yrinv$) vs. stellar mass, for central galaxies (solid) and satellites in bins of host halo mass (dashed and dot-dashed).
Both central and satellite quenched fractions increase strongly with stellar mass.
\textbf{(b)}: Same, but as a function of host halo mass split into bins of stellar mass.
Curves show satellites while points show central galaxies at their median host halo mass.
The satellite quenched fraction increases with halo mass and is always higher than the central galaxy quenched fraction at the same stellar mass.
\textbf{(c)}: Satellite quenched fraction excess, given by equation (\ref{eq:qufrac_excess}), vs. host halo mass.
This excess is nearly independent of satellite mass.
Error bars show 68\% confidence interval for a beta distribution \citep[see][]{Cam11}, and curves are offset slightly for clarity.
} \label{fig:qufrac-m}
\end{figure}

We now explore the galaxy and host halo mass dependence of the quenched fraction in detail.
Fig.~\ref{fig:qufrac-m}a shows that the likelihood of central galaxies to be quenched increases strongly with stellar mass, rising from 20\% to almost 95\% across a factor of 15 in stellar mass.
This mass range represents the strongest transition in galaxy star formation properties \citep[e.g.,][]{KauHecWhi03}.
By contrast, the satellite quenched fractions rise more gradually with stellar mass at fixed halo mass.
This is because low-mass satellites are more likely to be quenched, leaving less room for stellar mass dependence.

Fig.~\ref{fig:qufrac-m}b shows the same but demonstrates more directly how the quenched fraction increases with halo mass at fixed stellar mass.
Square points show central galaxies plotted at their median host halo mass for each stellar mass bin, while curves show satellites.
Satellites reside in higher mass host halos than central galaxies of the same stellar mass because, by definition, satellites are not the most massive galaxy in their group, so the total stellar mass, and hence halo mass, of a satellite's group is typically higher than if it were a (most massive) central galaxy.
At a given stellar mass, satellites are always more likely to be quenched than central galaxies.
This is true even in the lowest halo mass scales we probe ($3 \times 10^{11}\msol$), again implying no minimum halo mass for satellite quenching.
Comparing instead satellites to central galaxies in the same halo masses, central galaxies are always more likely to be quenched than the satellites in their same host halo.
This latter trend arises from the strong dependence of the quenched fraction on stellar mass.

Fig.~\ref{fig:qufrac-m}b shows that, at fixed stellar mass, transitioning from a central to a satellite galaxy in a progressively more massive host halo results in a significant increase in likelihood of quenching, but this increase with halo mass is weaker for more massive galaxies \citep[see][for a similar trend for galaxy age]{PasGalFon10}.
At face value, this might seem to imply that more massive galaxies are less affected by satellite-specific processes.
However, one must take care in interpreting \textit{absolute} changes in quenched fraction.
At lower stellar mass, central galaxies are less likely to be quenched and thus are able to experience a larger absolute change in quenched fraction after infall.
At higher stellar mass, there is less room for additional satellite-specific quenching.

We can examine host halo mass dependence more directly by removing the intrinsic dependence of the quenched fraction on stellar mass.
Specifically, we examine the \textit{relative} change in quenched fraction between central and satellite galaxies of the same stellar mass, and we scale this by the fraction of active (able to be quenched) central galaxies:
\begin{equation} \label{eq:qufrac_excess}
\fsatqexcess \equiv \frac{\fsatq - \fcenq}{\fcena} = 1 - \frac{\fsata}{\fcena}\, ,
\end{equation}
where $f_Q$ and $f_A$ are the fraction of quenched and active galaxies, respectively, related via $f_Q = 1-f_A$ and measured independently for central and satellite galaxies.
A similar statistic was used in \citet{vdBAquYan08} and \citet{TinWet10}, and more recently by \citet{PenLilRen11}, but our interpretation of it differs from those.
If the central galaxy active fraction has not evolved since the time of satellite infall, such that $\fcena(\zinf) = \fcena(z \approx 0)$, then $\fsatqexcess$ represents the fraction of satellites that quenched after infall.
However, as we will show in \citetalias{WetTinCon12a}, the typical redshift of infall for current satellites in our mass range is $z \approx 0.5$, so the assumption of no central galaxy evolution since the time infall is not a good one \citep[see also][]{TinWet10}.
To the extent that the central galaxy quenched fraction has increased over time, $\fsatqexcess$ more accurately represents the \textit{excess} fraction of satellites that quenched after infall that would not have quenched had they remained central galaxies, as can be seen by the right-hand side of equation (\ref{eq:qufrac_excess}).

Fig.~\ref{fig:qufrac-m}c shows that $\fsatqexcess$ depends only weakly on stellar mass at low halo mass, and it is remarkably \textit{independent} of stellar mass in halos $\gtrsim 3 \times 10^{13}\msol$.
(While the error at high stellar mass is large, the trend is consistent with that of lower stellar mass.)
Moreover, the dependence of $\fsatqexcess$ on stellar mass in lower mass halos does not necessarily imply a change in satellite quenching efficiency, for two reasons.
First, satellite dynamical infall times are shortest when the satellite-to-halo mass ratio is closest, being shorter than a Hubble time when the satellite halo to host halo mass ratio is closer than $\sim0.1$ \citep{BoyMaQua08, JiaJinFal08, WetWhi10}.
Thus, surviving satellites in this regime have necessarily fallen in more recently and have not had as much time to quench.
Second, purity and completeness in the group catalog are worse when the satellite-to-halo mass ratio is high, so satellite quenched fractions are most underestimated in this regime.

These satellite quenched fraction excess results show that \textit{the apparent reduction of the dependence of the quenched fraction on host halo mass at higher stellar mass primarily reflects that a higher fraction of more massive satellites quenched prior to infall.}

%% RADIUS %%%%%%%%%%%%%%%%%%%%%%%%%%%%%%%%%%%%%%%%%%%%%%%%%%%%%%%%%%%%%%%%%%%%%%%%%%%%%%%%
\section{Radial dependence} \label{sec:radius}

\subsection{Quenched fraction within $\rvir$}

Having explored the mass dependencies of the SSFR distribution and quenched fraction, we next examine how these quantities depend on location within the host halo.
Examining the dependence on halo-centric radius is particularly informative: because dynamical friction causes a correlation between satellite radius and time since infall (\citealt{GaoWhiJen04, WeivdBPas11}, also \citetalias{WetTinCon12b}), radial gradients translate into an evolutionary sequence of star formation.

In examining satellite radial gradients, one must ensure that the results are not affected by possible mass segregation (more massive satellite preferentially being at smaller radii).
We find no evidence for satellite mass segregation at any halo mass in our group catalog, in agreement with previous works \citep{PraDri05, HudSteSmi10, vdLWilKau10}, but see also \citet{vdBPasYan08}b.
Furthermore, we examine satellites in narrow bins of stellar mass, thus minimizing the effects of any possible mass segregation.

\begin{figure}
\centering
\includegraphics[width=0.99\columnwidth]{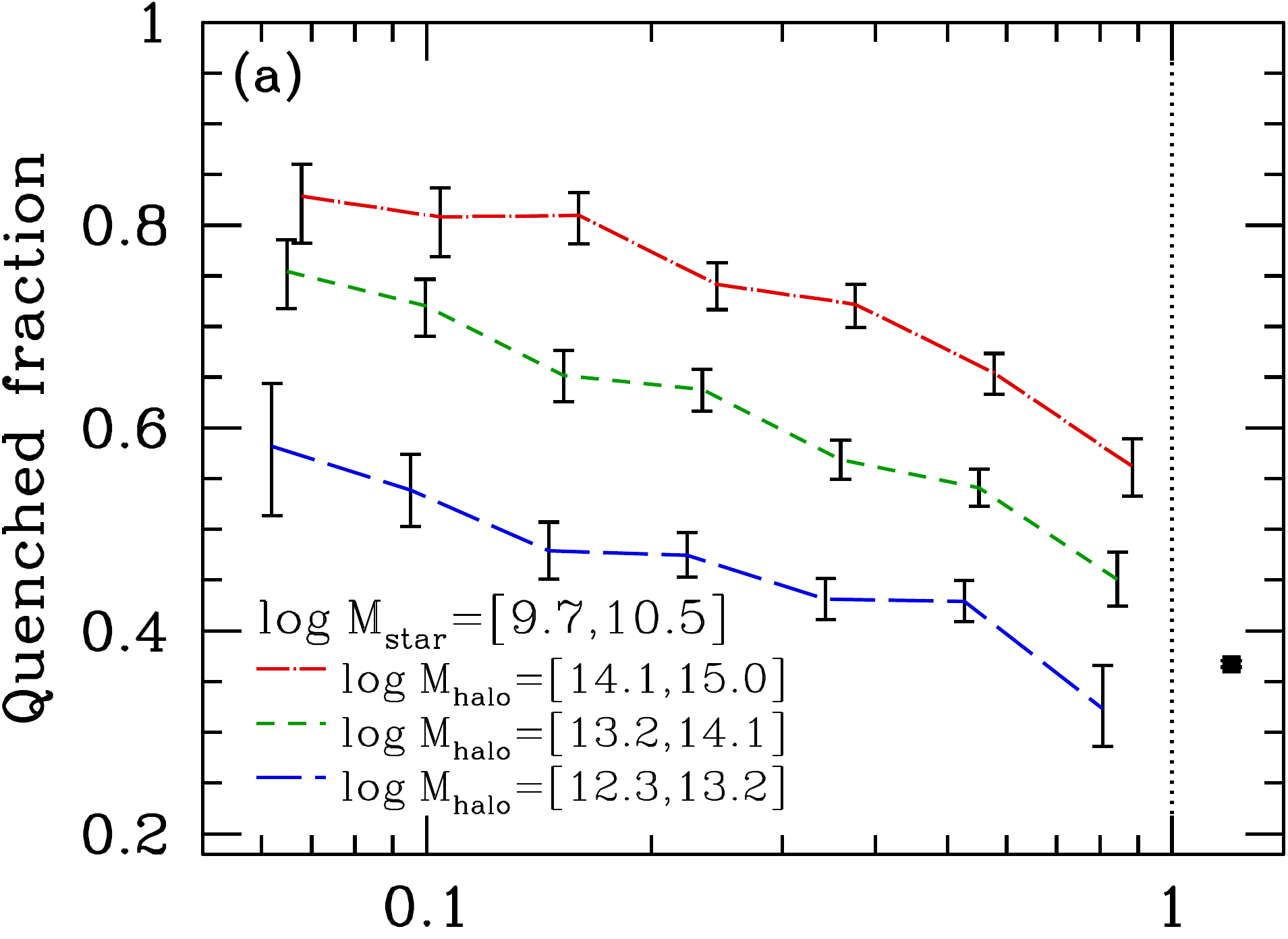}
\includegraphics[width=0.99\columnwidth]{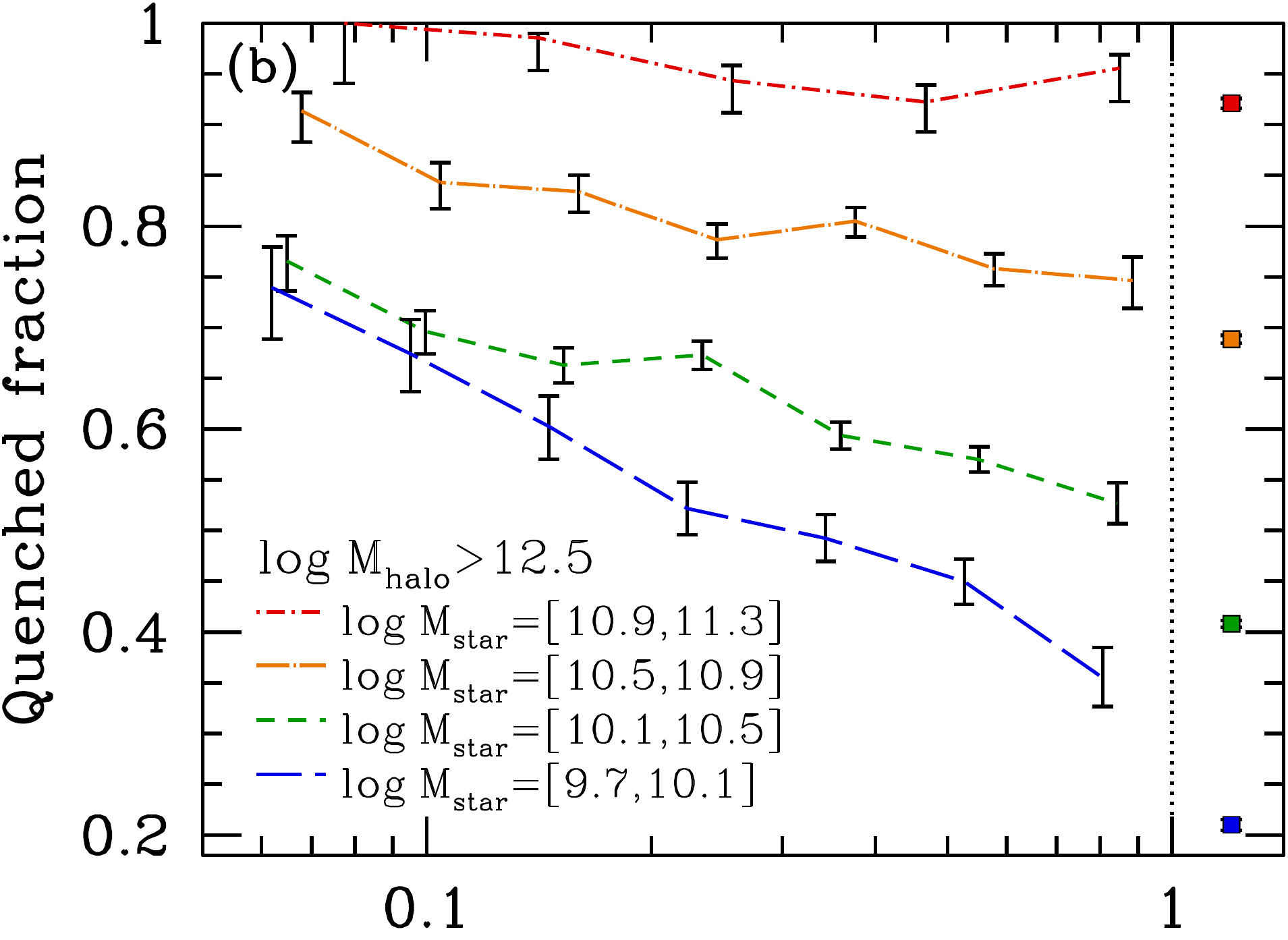}
\includegraphics[width=0.99\columnwidth]{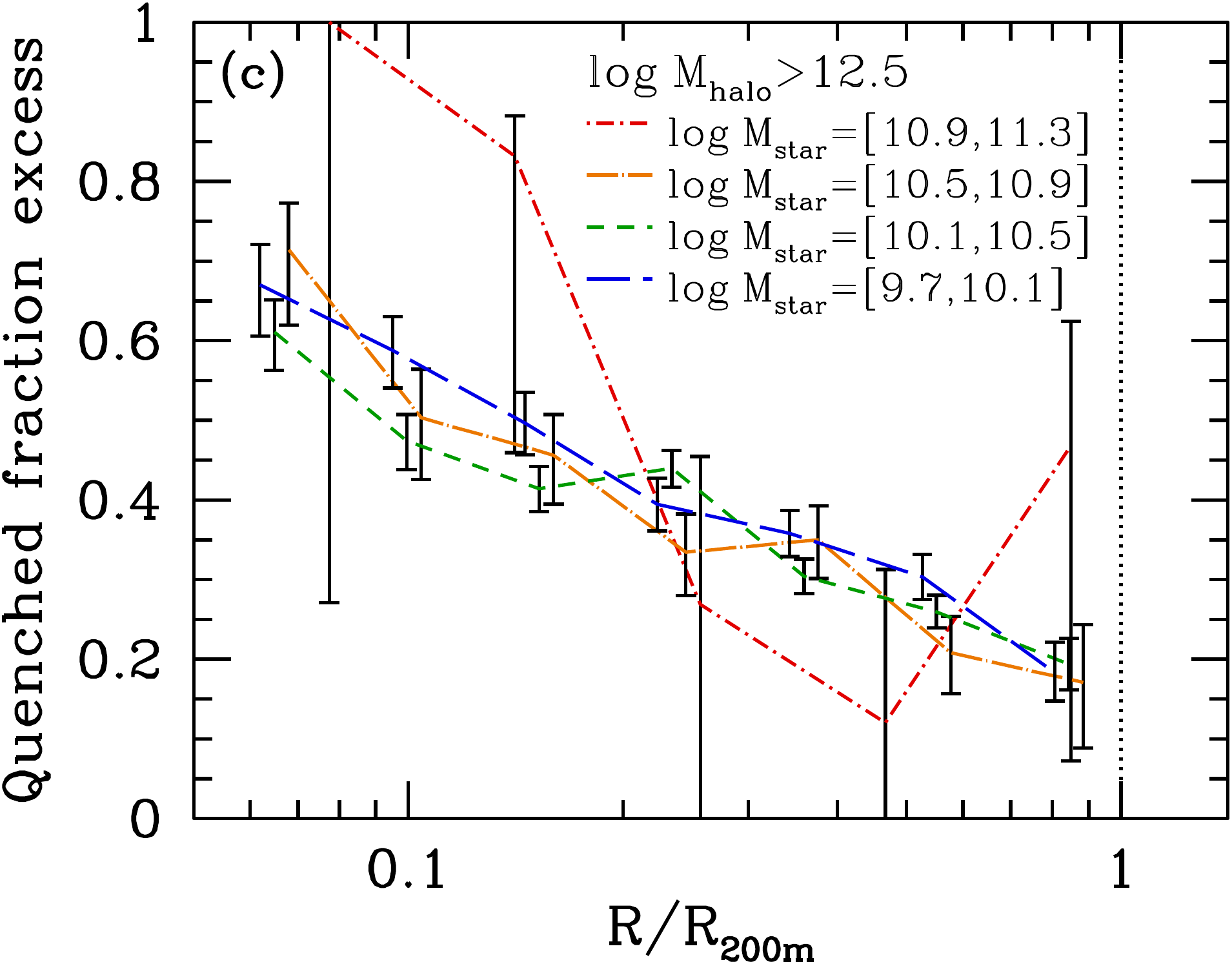}
\caption{
Galaxy quenched fraction vs. projected radius, scaled to the host halo virial radius.
Curves show satellites and points show central galaxies, regardless of location, of the same stellar mass.
\textbf{(a) and (b)}: Varying halo mass and satellite mass, respectively.
The quenched fraction increases with decreasing radius, and at all radii the satellite quenched fraction is higher in more massive halos and for more massive galaxies.
\textbf{(c)}: Satellite quenched fraction excess, given by equation (\ref{eq:qufrac_excess}), is independent of satellite mass at all radii.
Curves are offset slightly for clarity, and the highest stellar mass has reduced radial binning given limited statistics.
} \label{fig:qufrac-rad}
\end{figure}

Fig.~\ref{fig:qufrac-rad}a and b show the satellite quenched fraction as a function of projected radius scaled to the host halo virial radius, varying both halos mass and satellite mass, respectively.
For comparison, points show the average quenched fraction for all central galaxies of the same stellar mass.
At all satellite and halo masses, the quenched fraction increases toward smaller radius.
These gradients are strong, particularly for lower mass galaxies, which exhibit a 4 times higher quenched fraction at $0.05\,\rvir$ than in the field.
These gradients imply that a satellite's quenching likelihood depends on its time since infall and/or whether it passed through denser regions of its host halo.

Interestingly, the quenched fraction is higher in more massive halos even at fixed $R / \rvir$, by an amount that is independent of radius.
This result implies that \textit{satellite quenching does not simply depend on its currently observed local density, because at a given $R / \rvir$, halos of any mass have the same density} (modulo minor dependence on halo concentration).
This does not rule out the possibility that satellite quenching could occur via crossing above some fixed background density threshold, but it means that orbital evolution as/after quenching occurs changes the observed correlation with density.
In other words, currently observed local density is not necessarily indicative of the maximum local density a galaxy has ever experienced.
Thus, understanding satellite quenching requires understanding the detailed orbital histories of satellites, as we will examine in \citetalias{WetTinCon12b}.

The results of Fig.~\ref{fig:qufrac-rad}b show that radial gradients are shallower at higher stellar mass, which again might seem to imply that satellite quenching efficiency decreases with satellite mass.
To examine this more carefully, we again measure the satellite quenched fraction excess, given by equation (\ref{eq:qufrac_excess}), in bins of scaled radius.
Fig.~\ref{fig:qufrac-rad}c shows that, remarkably, $\fsatqexcess$ is \textit{independent} of satellite mass at all radii.
(Again, the error bars at high stellar mass are large, but the trend is consistent.)
This result strongly supports our conclusion that the apparent reduced dependence on local density at higher galaxy mass simply reflects a higher fraction of satellites having quenched prior to infall.

Note that the results in Fig.~\ref{fig:qufrac-rad} are for projected radius and incorporate purity and completeness effects.
At small projected radii the primary contamination is the incorporation of satellites in the same halo but at larger (line-of-sight) radii, while at large radii the primary contamination is from other central galaxies scattering into the halo.
Both effects lead to an underestimation of the true satellite quenched fraction at a given radius, so the true 3-D radial gradients could be somewhat steeper or shallower than in Fig.~\ref{fig:qufrac-rad}.

\subsection{Quenched fraction outside $\rvir$}

The results from Fig.~\ref{fig:qufrac-rad} show that the quenched fraction for satellites can be significantly higher than that of central galaxies even for satellites near the virial radius in high-mass halos.
This effect may be driven in part by these satellites having passed within much smaller radii, given that many satellites are on highly elliptical orbits \citep{Ben05, KhoBur06, Wet11}, and that galaxy groups falling into more massive halos can bring in already quenched satellites \citep{BerSteBul09, LiYeeEll09, McGBalBow09}.
However, it is interesting to consider whether environment affects star formation in galaxies beyond the virial radius by examining whether the radial gradients of Fig.~\ref{fig:qufrac-rad} persist to larger radii.
An enhanced red/quenched fraction has been noted in examining galaxies out to $2-4\,\rvir$ around massive clusters \citep{HanSheWec09, vdLWilKau10}.

\begin{figure*}
\centering
\includegraphics[width=2\columnwidth]{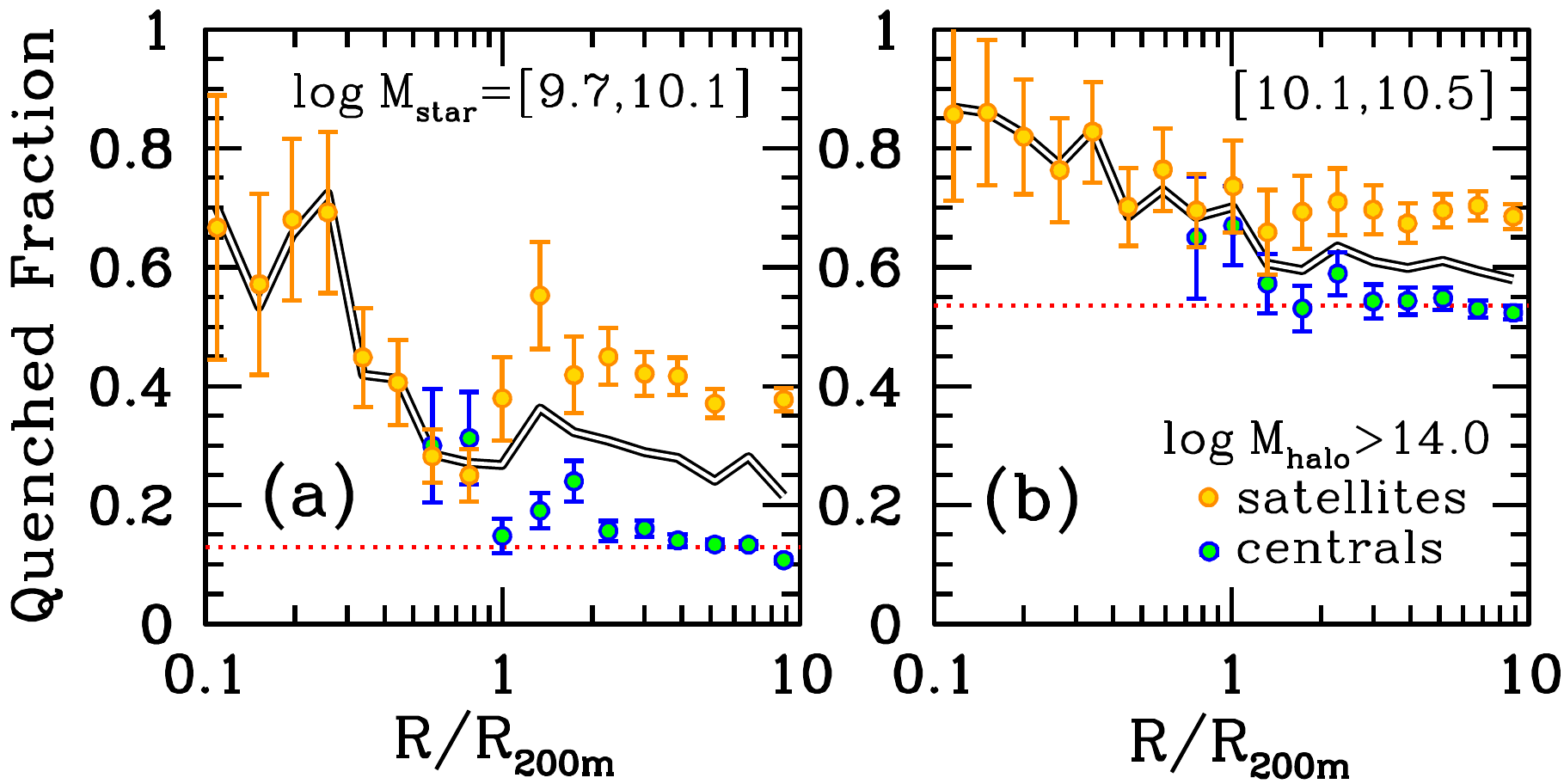}
\caption{
Galaxy quenched fraction vs. projected radius, scaled to the host halo virial radius, including \textit{all} galaxies within $10\,\rthm$ (projected) regardless of group assignment.
Panels show different stellar mass bins around cluster-mass halos ($\mthm > 10^{14}\msol$).
Solid curves show all galaxies, points show galaxies split into satellites and centrals, and dotted curve shows the average value for central galaxies at $5 - 10\,\rthm$.
Central galaxies deviate significantly from the field value only within $\sim2\,\rthm$.
} \label{fig:qufrac-rad_outer}
\end{figure*}

Fig.~\ref{fig:qufrac-rad_outer} shows the galaxy quenched fraction as a function of projected radius, similar to Fig.~\ref{fig:qufrac-rad}, except including \textit{all} galaxies (centrals and satellites) within a projected radius of $10\,\rvir$ around massive clusters, regardless of whether they are members of the selected halos or reside in different halos.
(Because we impose no redshift cut here, some galaxies within $\rvir$ are not members of the selected cluster-mass halos.)
The quenched fraction for all galaxies (solid curve) exhibits a clear enhancement out to $\sim3 - 4\,\rvir$.
However, if we decompose this population into central and satellite galaxies according to the group catalog, it is clear that the enhancement beyond $\rvir$ is driven primarily by satellites in different halos and a declining ratio of satellites-to-centrals as a function of radius from the clusters.
The latter simply reflects an increased likelihood of finding massive groups closer to clusters.
Indeed, compared with the central galaxy quenched fraction measured at $5-10\,\rvir$ (dotted curve), central galaxies deviate significantly from the field value only within $\sim2\,\rvir$.
At these radii, the enhancement is likely caused by (1) massive halos being highly ellipsoidal, and (2) satellites on orbits with apocenters beyond the virial radius \citep{GilKneGib05, LudNavSpr09, WanMoJin09}.
We explore the latter explanation in more detail in \citetalias{WetTinCon12b}.
See also Y. \citet{WanYanMo09} for similar trends for much fainter galaxies ($M_r < -17$).

Thus, by decomposing galaxies beyond $\rvir$ into centrals and satellites, we find no compelling evidence that central galaxy star formation is affected significantly prior to infall.
Instead, the SSFR gradients beyond the virial radius are caused primarily by satellites in groups and that massive groups are more likely to be found closer to clusters.
This agrees with our results in \citetalias{TinWetCon11}, where we showed that the change in the overall galaxy quenched fraction with large-scale ($10\hmpc$) environmental density is caused simply by a changing halo mass function.
Thus, \textit{all environmental effects on galaxy star formation are consistent with being governed only by a galaxy's host halo.}

\subsection{Star formation rate distribution}

\begin{figure}
\centering
\includegraphics[width=0.99\columnwidth]{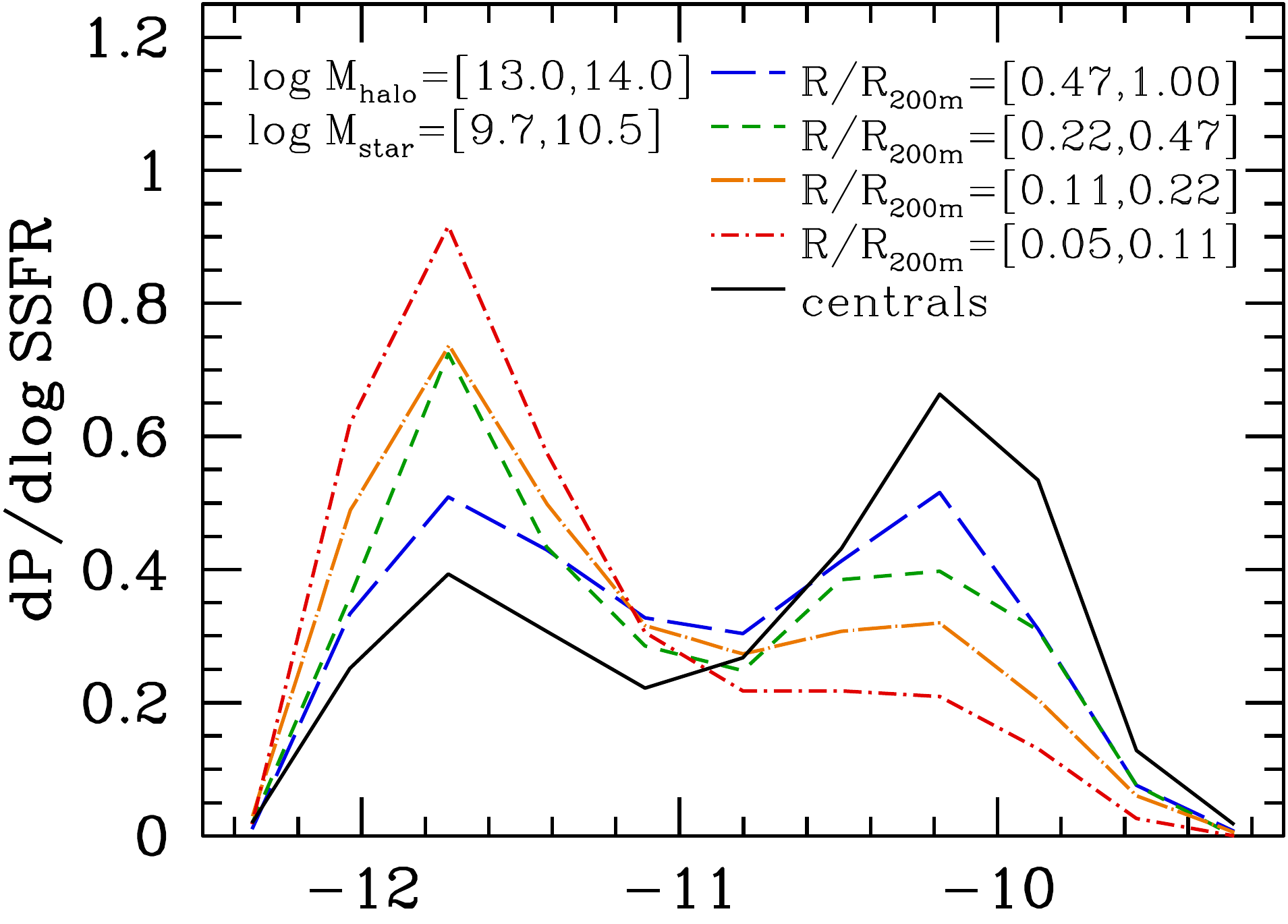}
\includegraphics[width=0.99\columnwidth]{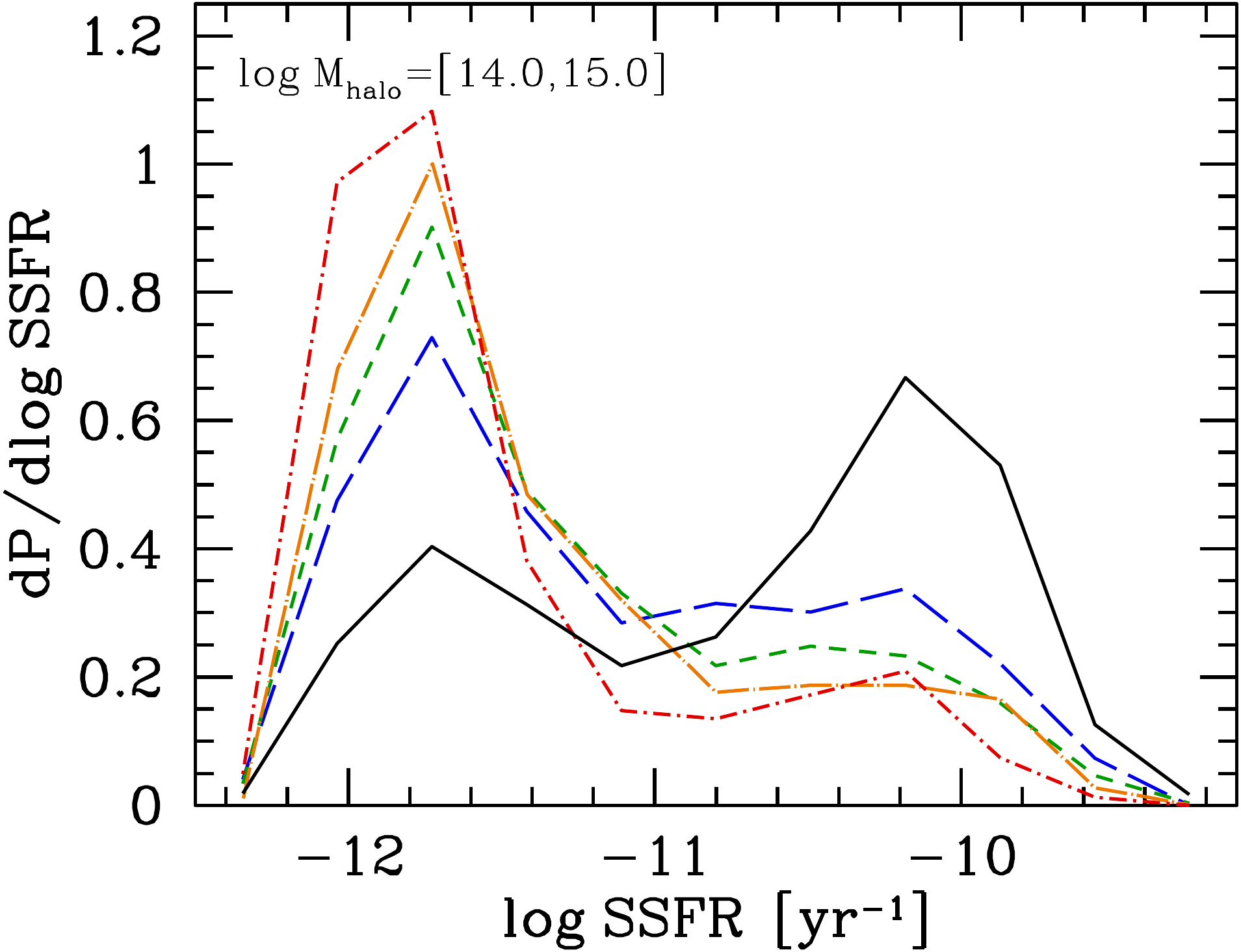}
\caption{
SSFR distribution for central galaxies (solid) and satellites in bins of projected radius, scaled to the host halo virial radius (dashed and dot-dashed).
Panels show different host halo mass bins.
The SSFR bimodality persists at all radii, with a constant location for the active galaxy peak, bimodality break, and fraction of galaxies at the break.
The only difference is a decreased fraction of active galaxies at smaller radii.
} \label{fig:ssfr_distr-rad}
\end{figure}

Finally, to understand more fully the radial gradients in Fig.~\ref{fig:qufrac-rad}, we examine how the full SSFR distribution changes with halo-centric radius.
Fig.~\ref{fig:ssfr_distr-rad} shows the satellite SSFR distribution in bins of scaled radius, with panels showing different host halo masses.
(We show low-mass satellites, for which changes in the SSFR distribution are most clearly visible, though similar trends persist for higher mass satellites.)
For comparison, the solid curves show central galaxies of the same stellar mass.
Remarkably, the bimodal nature of the SSFR distribution persists at all radii within all halo masses we probe.
Similar to the halo mass trends in Fig.~\ref{fig:ssfr_peak-mhalo}, there is no significant change in the active galaxy SSFR peak location with radius at the $\sim0.1$ dex level, nor is there a significant change in the location of the bimodality break or the fraction of galaxies there.
Thus, \textit{star formation in currently active satellites is not affected by local density or time since infall.}
We now discuss the strong implications of this persistent SSFR bimodality.

%% BIMODALITY %%%%%%%%%%%%%%%%%%%%%%%%%%%%%%%%%%%%%%%%%%%%%%%%%%%%%%%%%%%%%%%%%%%%%%%%%%%%
\section{Implications of the star formation rate bimodality} \label{sec:bimodality}

We have shown that all galaxies, central and satellite, exhibit a similar bimodal distribution of SSFRs, with a deficit at $\ssfr \approx 10^{-11}\yrinv$ and the same active galaxy SSFR peak at fixed stellar mass.
This bimodality extends across all stellar masses, host halo masses, and halo-centric radii; in no regime is there \textit{ever} a pile-up of galaxies at intermediate SSFRs in the `green valley'.
Thus, halo dependence of satellite galaxy SFR manifests itself simply as a decrease/increase in the fraction of active/quenched satellites with increasing halo mass and with decreasing halo-centric radius.

This `universal' SSFR bimodality places strong constraints on satellite-specific quenching mechanisms and timescales.
In particular, to the extent that halo-centric radius selection represents an evolutionary sequence, the results of Figs.~\ref{fig:ssfr_distr-m} and \ref{fig:ssfr_distr-rad} have three clear implications.
First, the fact that active satellites have the same SSFR peak and distribution as active central galaxies, even within $0.1\,\rvir$, implies that a significant fraction of satellites have evolved in the same manner as central galaxies, unaffected by host halo processes after infall.
This suggests that any halo-specific quenching process takes significant time (at least an orbital time of several Gyrs) to \textit{start} to affect satellite star formation \citep[see also][]{McGBalWil11}.

Second, the persistent bimodality break implies that once quenching begins, the transition to quenched SSFRs \textit{always} occurs on a rapid timescale.
Specifically, once begun, the fading timescale must be significantly shorter than the timescale over which satellites remain active after infall (several Gyrs), and likely shorter than molecular gas consumption timescale ($\sim2.4\gyr$, e.g., \citealt{BigLerWal11}), because if satellite SFR simply were to fade slowly over this timescale after infall, we would observe an excess of galaxies at intermediate SSFR values in the `green valley'.
However, the fact that the fraction of satellites at $\ssfr \approx 10^{-11}\yrinv$ is both non-zero and constant implies that satellite SFR fading does take cosmologically non-trivial time to occur, and that this fading timescale does not depend on host halo mass or halo-centric radius.
It remains unclear if this scenario can be achieved merely by a sharp cold gas density threshold for sustaining star formation \citep[see e.g.,][]{KruMcKTum09}, or if some additional process is needed to accelerate cold gas consumption and SFR decay.
We will return to this question in more detail in \citetalias{WetTinCon12a}.

Third, the deviations of the satellite SSFR distribution from that of central galaxies of the same stellar mass down to the lowest halo mass scales we probe ($3 \times 10^{11}\msol$) means that we find no minimum halo mass for satellite-specific processes.
This result is consistent with \citet{TolBoyBar11}, who found an enhanced red fraction of Large Magellanic Cloud-mass galaxies if they are near Milky Way-mass galaxies in SDSS.
Furthermore, the increasing quenched fraction with halo mass implies either that satellite quenching is more efficient and begins more quickly in higher mass halos, or simply that satellites in more massive halos have preferentially been satellites for longer, possibly from satellite pre-processing in groups.
In \citetalias{WetTinCon12a} and \citetalias{WetTinCon12b}, we will demonstrate that the latter explanation is more likely.

Thus, the persistent bimodality implies that a delayed-then-rapid quenching mechanism must drive satellite quenching.
While the implied long timescale for satellite quenching has led several authors to propose strangulation as the dominant mechanism \citep[e.g.,][]{BalNavMor00, McGBalBow09, WeiKauvdL10}, it is unclear that a mode of strangulation in which star formation fades slowly upon infall can retain the SSFR bimodality at all radii without putting too many galaxies at intermediate SSFR (though see \citealt{WeiKauvdL10}, and also \citep{BalMcGWil09}).
One possibility is that, while the satellite's hot gas halo is being stripped from the outside, the innermost hot gas remains unaffected such that cooling and star formation continue as normal for several Gyrs.
This idea is consistent with the X-ray observations of \citet{SunJonFor07} and \citet{JelBinMul08}, who found that more than half of bright satellites have significantly truncated yet detectable hot gas halos.

A natural mechanism to rapidly quench star formation on a short timescale ($\lesssim 500\myr$) is ram-pressure stripping, which can remove cold gas directly from a satellite galaxy's disc.
However, while ram-pressure is observed to act on galaxies in high-mass clusters, it has not been observed in lower mass groups, where lower halo gas densities and satellite velocities likely lower its efficiency.
Moreover halos of $\mvir \lesssim 10^{12}\msol$ are not expected to have virial shock fronts which support hot, virialized gas within the halo \citep{DekBir06}, so in this mass regime it is not clear that either strangulation or ram-pressure can be efficient.
This may suggest the need for tidal stripping, or for harassment and/or mergers induce rapid cold gas consumption that quenches star formation on short timescales.
In \citetalias{WetTinCon12b}, we will test quantitatively whether these mechanisms can reproduce the galaxy mass, halo mass, and halo-centric radius dependencies that we have outlined here.

%% DISCUSSION %%%%%%%%%%%%%%%%%%%%%%%%%%%%%%%%%%%%%%%%%%%%%%%%%%%%%%%%%%%%%%%%%%%%%%%%%%%%
\section{Summary and Conclusion} \label{sec:summary}

Using galaxy group catalogs created from SDSS Data Release 7, we examined in detail the SSFRs of satellite galaxies and how they depend on galaxy mass, host halo mass, and halo-centric radius.
Our galaxy sample and group catalog provide good statistics across of wide range of masses both for satellites ($\mstel = 5 \times 10^{9} - 2 \times 10^{11}\msol$) and their host halos ($\mthm = 3 \times 10^{11} - 10^{15}\msol$), and our use of spectroscopically determined SSFR means our results are insensitive to dust reddening, important for low-mass galaxies
Our main results are as follows.

\textbf{Persistent Bimodality:}
All galaxies that we probe, regardless of galaxy mass, host halo mass, and halo-centric radius, exhibit similar bimodal SSFR distributions, including a fixed break at $\ssfr = 10^{-11}\yrinv$, the same fraction of galaxies at the break, and the same SSFR peak for active galaxies.
Satellites are simply more likely to lie on the quenched side of the distribution.
This persistent bimodality implies that a significant fraction of satellites have evolved in the same way as central galaxies, independent of their host halo, for several Gyrs.
Once begun, the satellite SFR fading process is rapid as compared with time over which satellites remain active after infall and thus is likely shorter than a $\sim2.4\gyr$ cold gas consumption timescale, but it does not depend on host halo mass or halo-centric radius.

\textbf{Clear host halo mass and radius dependences within the virial radius:}
At fixed stellar mass, satellites are more likely to be quenched than central galaxies, even down to the lowest halo masses we probe ($3 \times 10^{11}\msol$).
This quenched likelihood increases at smaller halo-centric radius, further implying that satellite quenching takes several Gyrs to occur after infall.
The quenched likelihood also monotonically increases with halo mass across our entire halo mass range, even at fixed $R / \rvir$, which implies that satellite quenching does not simply depend on \textit{current} local density.
It is not immediately clear whether satellite quenching is more efficient/rapid in higher mass halos, or surviving satellites in that regime have simply been satellites longer.

\textbf{Unclear stellar mass dependence:}
The central galaxy quenched fraction increases significantly with stellar mass.
While more massive satellites show a weaker absolute change in quenched fraction with host halo mass or radius, this is largely a manifestation of central galaxy mass dependence prior to infall.
The satellite quenched fraction excess, which more directly measures quenching efficiency, shows little-to-no dependence on galaxy mass.

\textbf{Importance of satellite-specific evolution:}
In \citetalias{TinWetCon11}, we showed that the star formation histories and quenched fractions of central galaxies at fixed galaxy (or halo) mass are essentially independent of their large-scale ($10\hmpc$) environment.
Here, we have shown that this lack of environmental dependence extends to central galaxies as close as $\sim2\,\rvir$ to massive clusters.
Together, these results demonstrate the importance of satellites in understanding galaxy evolution, because satellite-specific processes are the only significant environmental processes that affect galaxy star formation and color.

The results of the satellite quenched fraction excess suggest that infalling active satellites quench at least as efficiently/rapidly at high stellar mass as at low stellar mass.
However, it remains true that satellite-specific processes are less important for building up the red sequence at higher stellar mass, because more galaxies quench as central galaxies prior to infall in that regime \citep[see also][]{vdBAquYan08}.
Unfortunately, it is difficult to determine directly from $\fsatqexcess$ whether satellite quenching is fully independent of galaxy mass because doing so requires knowing in detail the quenched fraction as a function of stellar mass prior to satellite infall.
In \citetalias{WetTinCon12a}, we will use a high-resolution cosmological simulation to obtain the infall time distributions of satellites, combined with observational constraints on high-redshift quenched fractions, to examine this issue more carefully.

The dependences on mass and halo-centric radius that we see broadly agree with many previous works using SDSS catalogs.
This includes a monotonic increase in quenched fraction with halo mass mass which nearly equals the intrinsic dependence on galaxy mass \citep{WeivdBYan06a, BlaBer07, HanSheWec09, KimSomYi09}, as well as satellites being more likely to be quenched at smaller radii \citep{WeivdBYan06a, BlaBer07, HanSheWec09} with less massive satellites exhibiting stronger radial gradients \citep{vdLWilKau10}.
Our catalog enabled us to extend many of these results to a wider range of galaxy and halo masses, particularly at the low mass end, and our use of spectroscopically-derived SSFR ensures that of quenched fraction results are not biased by dust in that regime.
Furthermore, our SSFR bimodality results qualitatively agree with previous results on the color bimodality as a function of projected galaxy density \citep{BalBalNic04, BalBalBow06} and the more recent results of \citet{McGBalWil11} and \citet{PenLilRen11}, who showed similar SSFR distributions and active galaxy SSFR values when splitting all galaxies into groups vs. the field.
We have shown that this `universal' SSFR bimodality extends to specifically satellite galaxies in all halo masses and at all radii.

Using color cuts to measure red galaxy fractions, \citet{vdBAquYan08}, and more recently \citet{PenLilRen11}, employed a similar statistic as our satellite quenched fraction excess to show that this excess is independent of satellite mass.
We have shown that this results extends to more robust measures of SFR, and to all halo masses and at all radii.
However, our physical interpretation of this statistic differs from theirs.
In particular, this statistic only measures the \textit{excess} fraction of satellites that quenched after infall that would not have quenched had they remained central galaxies, because the $z \approx 0$ central galaxy quenched fraction does not represent the initial conditions of satellites at the time of infall.
We will explore this issue further in \citetalias{WetTinCon12a}.

The mass and radius dependences we see disagree with those of \citet{vdBPasYan08}b, who found only weak dependence of galaxy properties on halo mass and radius at fixed stellar mass.
However, they examined primarily mean $g-r$ color of satellites as a measure of star formation, which is problematic because dust reddening causes a severe biasing of $g-r$ color, particularly for low-mass galaxies, and the mean value is a less sensitive characterization of a highly bimodal distribution.
Indeed, we find similar results to theirs when examining mean $g-r$ color.

The satellite SSFR distribution dependencies we have detailed in this paper provide tight constrains on models for satellite-specific evolution.
In \citetalias{WetTinCon12a}, we will use a high-resolution cosmological simulation to obtain the detailed orbital histories of satellites, which we will use to more rigorously constrain the timescales of satellite quenching, and in \citetalias{WetTinCon12b} we will develop models for satellite-specific quenching via different mechanisms to test which reproduce the trends observed in this paper.

%% ACKNOWLEDGE %%%%%%%%%%%%%%%%%%%%%%%%%%%%%%%%%%%%%%%%%%%%%%%%%%%%%%%%%%%%%%%%%%%%%%%%%%%
\section*{Acknowledgments}

A.W. acknowledges partial support through an NSF Graduate Research Fellowship and thanks the organizers, participants, and staff of the \textit{Monsters, Inc.} workshop at the Kavli Institute for Theoretical Physics, supported in part by the NSF, where some of this work was completed.
We thank Michael Blanton, David Hogg, and collaborators for publicly releasing the NYU VAGC, as well as Jarle Brinchmann and the MPA-JHU collaboration for publicly releasing their spectral reductions.
We thank Frank van den Bosch and Jarle Brinchmann for insightful conversations, as well as Frank van den Bosch and Martin White for comments on an early draft.

Funding for the SDSS and SDSS-II has been provided by the Alfred P. Sloan Foundation, the Participating Institutions, the National Science Foundation, the U.S. Department of Energy, the National Aeronautics and Space Administration, the Japanese Monbukagakusho, the Max Planck Society, and the Higher Education Funding Council for England.
The SDSS Web Site is http://www.sdss.org/.
The SDSS is managed by the Astrophysical Research Consortium for the Participating Institutions.
The Participating Institutions are the American Museum of Natural History, Astrophysical Institute Potsdam, University of Basel, University of Cambridge, Case Western Reserve University, University of Chicago, Drexel University, Fermilab, the Institute for Advanced Study, the Japan Participation Group, Johns Hopkins University, the Joint Institute for Nuclear Astrophysics, the Kavli Institute for Particle Astrophysics and Cosmology, the Korean Scientist Group, the Chinese Academy of Sciences (LAMOST), Los Alamos National Laboratory, the Max-Planck-Institute for Astronomy (MPIA), the Max-Planck-Institute for Astrophysics (MPA), New Mexico State University, Ohio State University, University of Pittsburgh, University of Portsmouth, Princeton University, the United States Naval Observatory, and the University of Washington.

%\bibliography{biblio}
\bibliography{ms}

%% APPENDIX %%%%%%%%%%%%%%%%%%%%%%%%%%%%%%%%%%%%%%%%%%%%%%%%%%%%%%%%%%%%%%%%%%%%%%%%%%%%%%
%\appendix

\label{lastpage}

\end{document}